\documentclass[a4paper]{jpconf}
\usepackage{graphicx}
\usepackage{amssymb,amsmath,cite}

\newcommand{\ba}{\begin{eqnarray}}
\newcommand{\ea}{\end{eqnarray}}
\newcommand{\ban}{\begin{eqnarray*}}
\newcommand{\ean}{\end{eqnarray*}}
\newcommand{\bsub}{\begin{subequations}}
\newcommand{\esub}{\end{subequations}}

\def\ket#1{|#1\rangle}
	
\def\bsu3{\overline{{\rm SU(3)}}}
\def\bso6{\overline{{\rm SO(6)}}}
\def\bPi2{\overline{\Pi}^{(2)}}
\def\blam{\bar{\lambda}}
\def\bmu{\bar{\mu}}
\def\bK{\bar{K}}
\def\b0{\beta_0}
\def\beq{\beta_{\rm eq}}
\def\g0{\gamma_0}
\def\gaeq{\gamma_{\rm eq}}
\def\bs{\beta_*}

\begin{document}
\title{Coexisting partial dynamical symmetries\\
and multiple shapes}

\author{A Leviatan and N Gavrielov}

\address{Racah Institute of Physics, The Hebrew University, 
Jerusalem 91904, Israel}

\ead{ami@phys.huji.ac.il, noam.gavrielov@mail.huji.ac.il}

\begin{abstract}
We present an algebraic procedure for constructing Hamiltonians with several 
distinct partial dynamical symmetries (PDSs), of relevance to 
shape-coexistence phenomena. 
The procedure relies on a spectrum generating algebra 
encompassing several dynamical symmetry (DS) chains and a coherent state 
which assigns a particular shape to each chain.
The PDS Hamiltonian maintains the DS solvability and quantum numbers 
in selected bands, associated with each shape, and mixes other states. 
The procedure is demonstrated for a variety of multiple quadrupole shapes 
in the framework of the interacting boson model of~nuclei.
\end{abstract}
\section{Introduction}
\label{sec:intro}

During the past several decades, the concept of dynamical symmetry (DS) 
has become the cornerstone of algebraic modeling of dynamical 
systems~\cite{BNB,ibm,ibfm,vibron,LieAlg}. 
Its basic paradigm is a chain of nested algebras,
\ba
G_{\rm dyn} \supset G_1 \supset G_2 \supset \dots \supset G_{\rm sym} 
\;\;\;\qquad
\ket{\lambda_{\rm dyn},\, \lambda_1,\,\lambda_2,\,
\ldots,\,\lambda_{\rm sym}} ~,\quad 
\label{DS-chain}
\ea
and an Hamiltonian of the system written in terms of the Casimir 
operators, $\hat{C}[G]$,  of the algebras in the chain 
\ba
\hat{H}_{\rm DS} = \sum_{G} a_{G}\,\hat{C}[G] ~.
\label{H-DS}
\ea
In such a case, the spectrum is completely solvable, the energies and 
eigenstates are labeled by quantum numbers 
$(\lambda_{\rm dyn},\, \lambda_1,\,\lambda_2,\,\ldots,\,\lambda_{\rm sym})$, 
which are the labels of 
irreducible representations (irreps) of the algebras in the chain.
In Eq.~(\ref{DS-chain}), $G_{\rm dyn}$ is the dynamical (spectrum generating) 
algebra of the system such that operators of all physical observables can 
be written in terms of its generators 
and $G_{\rm sym}$ is the symmetry algebra. The DS spectrum 
exhibits a hierarchy of splitting but no mixing. 
A given $G_{\rm dyn}$ can encompass several DS chains, each providing 
characteristic analytic expressions for observables and definite 
selection rules.

A notable example of such algebraic setting 
is the interacting boson model 
(IBM)~\cite{ibm}, widely used in the description of low-lying quadrupole 
collective states in nuclei in terms of $N$ monopole $(s)$ and
quadrupole $(d)$ bosons, representing valence nucleon pairs.
The model is based on $G_{\rm dyn}\!=\!{\rm U(6)}$ and 
$G_{\rm sym}\!=\!{\rm SO(3)}$. The Hamiltonian is expanded in terms of the 
generators of U(6), $\{s^{\dag}s,\,s^{\dag}d_{m},\, d^{\dag}_{m}s,\, 
d^{\dag}_{m}d_{m '}\}$, 
and consists of Hermitian, rotational-scalar interactions 
which conserve the total number of $s$- and $d$- bosons, 
$\hat N \!=\! \hat{n}_s + \hat{n}_d \!=\! 
s^{\dagger}s + \sum_{m}d^{\dagger}_{m}d_{m}$. 
The solvable limits of the IBM correspond to the following DS chains
\bsub
\ba
&&{\rm U(6)\supset U(5)\supset SO(5)\supset SO(3)} \;\,\quad
\quad\ket{N,\, n_d,\,\tau,\,n_{\Delta},\,L} \quad 
{\rm spherical\; vibrator}~,\qquad
\label{U5-ds}
\\
&&{\rm U(6)\supset SU(3)\supset SO(3)} \;\;\;\qquad\quad\quad
\;\;\,\ket{N,\, (\lambda,\mu),\,K,\, L} \quad
\,{\rm prolate\!-\!deformed\;rotor}~,\qquad 
\label{SU3-ds}
\\
&&{\rm U(6)\supset \bsu3\supset SO(3)} \;\;\;\qquad\quad\quad
\;\;\,\ket{N,\, (\blam,\bmu),\,\bar{K},\, L} \quad
\,{\rm oblate\!-\!deformed\;rotor}~,\qquad 
\label{SU3bar-ds}
\\
&&
{\rm U(6)\supset SO(6)\supset SO(5)\supset SO(3)} \quad
\;\;\;\ket{N,\, \sigma,\,\tau,\,n_{\Delta},\, L} \quad 
\;{\rm \gamma\!-\!unstable\; deformed\; rotor}~.\qquad
\label{SO6-ds}
\ea
\label{IBMchains}
\esub
Here $N,n_d,(\lambda,\mu),(\blam,\bmu),\sigma,\tau,L$, 
label the relevant irreps of 
U(6), U(5), SU(3), $\bsu3$, SO(6), SO(5), SO(3), 
respectively, and $n_{\Delta},K,\bar{K}$ are multiplicity labels. 
The indicated basis states are eigenstates of the Casimir operators 
in the chain, with eigenvalues listed in Table~1 
for the leading sub-algebras $G_1$, and eigenvalues 
$L(L+1)$ [$\tau(\tau+3)$] for SO(3) [SO(5)]. 
The resulting DS spectra of the above chains resemble 
known paradigms of nuclear collective structure, 
as mentioned in Eq.~(\ref{IBMchains}), 
involving vibrations and rotations of a quadrupole shape. 
\begin{center}
\begin{table}[t]
\caption{\label{TabIBMcas}
\small
Eigenvalues of the Casimir operators, $\hat{C}_{k}[G]$ of order $k\!=\!1,2$, 
for the leading sub-algebras ($G_1$) of the DS-chains~(\ref{IBMchains}). 
The equilibrium deformations $(\beta_{\rm eq},\gamma_{\rm eq})$ 
define the quadrupole shape associated with each chain and determine 
the $G_1$-symmetry of 
$\ket{\beq,\gaeq;N}$, Eq.~(\ref{int-state}). The latter is an extremal state 
in a particular irrep ($\lambda\!=\!\Lambda_0$) of $G_1$, and serves as an 
intrinsic state for the respective ground-band.}
\vspace{1mm}
\centering
\begin{tabular}{llll}
\br
& & &\\[-3mm]
Algebra & Eigenvalues of &
Equilibrium deformations &
${\rm G_1}$-symmetry of $\ket{\beq,\gaeq;N}$\\
$\quad {\rm G_1}$ & $\quad\hat{C}_{k}[{\rm G_1}]$
& $\qquad(\beta_{\rm eq},\gamma_{\rm eq})$ 
& $\qquad\lambda_1=\Lambda_0$\\[4pt]
& & & \\[-3mm]
\mr
& & & \\[-2mm]
{\rm U(5)} & $n_d$ &  
$\,\beta_{\rm eq}\!=\!0$ & $\qquad n_d=0$ \\[2pt]
{\rm SU(3)}$\quad$ & $\lambda^2 +(\lambda+\mu)(\mu+3)\quad$ &
$(\beta_{\rm eq} \!=\!\sqrt{2},\gamma_{\rm eq}\!=\!0)$ & 
$\qquad (\lambda,\mu)=(2N,0)$\\[2pt]
$\bsu3$ & $\blam^2 +(\blam+\bmu)(\bmu+3)$ &
$(\beta_{\rm eq} \!=\!\sqrt{2},\gamma_{\rm eq}\!=\!\pi/3)$ &
$\qquad (\blam,\bmu)=(0,2N)$\\[2pt]
{\rm SO(6)} & $\sigma(\sigma+4)$ &
$(\beta_{\rm eq}\!=\!1,\gamma_{\rm eq}\,\,{\rm arbitrary})\quad$ &
$\qquad \sigma =N$\\[2pt]
& & &\\[-3mm]
\br
\end{tabular}
\label{Tab1}
\end{table}
\end{center}

\vspace{-18pt}
Geometry is introduced in the algebraic model by means of a coset space 
$U(6)/U(5)\otimes U(1)$  and a `projective' coherent 
state~\cite{gino80,diep80},
\bsub
\ba
\vert\beta,\gamma ; N \rangle &=&
(N!)^{-1/2}(b^{\dagger}_{c})^N\,\vert 0\,\rangle ~,\\
b^{\dagger}_{c} &=& (1+\beta^2)^{-1/2}[\beta\cos\gamma 
d^{\dagger}_{0} + \beta\sin{\gamma} 
( d^{\dagger}_{2} + d^{\dagger}_{-2})/\sqrt{2} + s^{\dagger}] ~,
\ea
\label{int-state}
\esub
from which an energy surface is derived by the 
expectation value of the Hamiltonian,
\ba
E_{N}(\beta,\gamma) &=& 
\langle \beta,\gamma; N\vert \hat{H} \vert \beta,\gamma ; N\rangle ~. 
\label{enesurf}
\ea 
Here $(\beta,\gamma)$ are
quadrupole shape parameters whose values, $(\beta_{\rm eq},\gamma_{\rm eq})$, 
at the global minimum of $E_{N}(\beta,\gamma)$ define the equilibrium 
shape for a given Hamiltonian. 
The shape can be spherical $(\beta =0)$ or 
deformed $(\beta >0)$ with $\gamma =0$ (prolate), $\gamma =\pi/3$ (oblate), 
$0 < \gamma < \pi/3$ (triaxial), or $\gamma$-independent. 
The coherent state with the equilibrium deformations, $\ket{\beq,\gaeq;N}$, 
serves as an intrinsic state for the ground band, whose rotational members 
are obtained by angular momentum projection. 
The equilibrium deformations associated with the 
DS limits, Eq.~(\ref{IBMchains}), are listed in Table~1 and 
conform with their geometric interpretation.
For these values, the ground-band intrinsic state, $\ket{\beq,\gaeq;N}$, 
becomes a lowest (or highest) weight state 
in a particular irrep ($\lambda_1=\Lambda_0$) of the leading sub-algebra 
$G_1$, as disclosed in Table~1.
\begin{figure}[t!]
\centering
\hspace{-0.7cm}
\includegraphics[width=5.6cm]{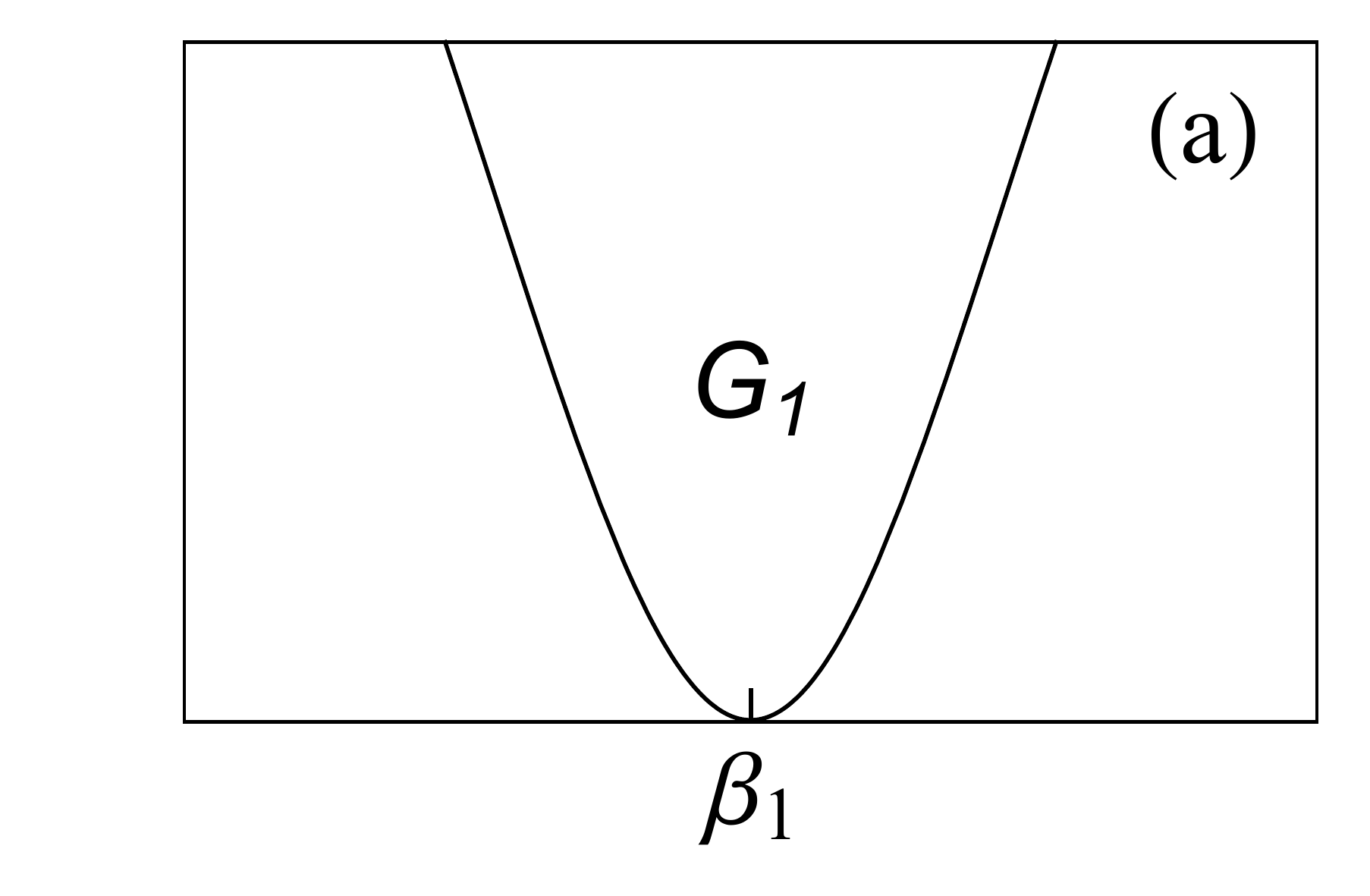}
\hspace{-0.4cm}
\includegraphics[width=5.6cm]{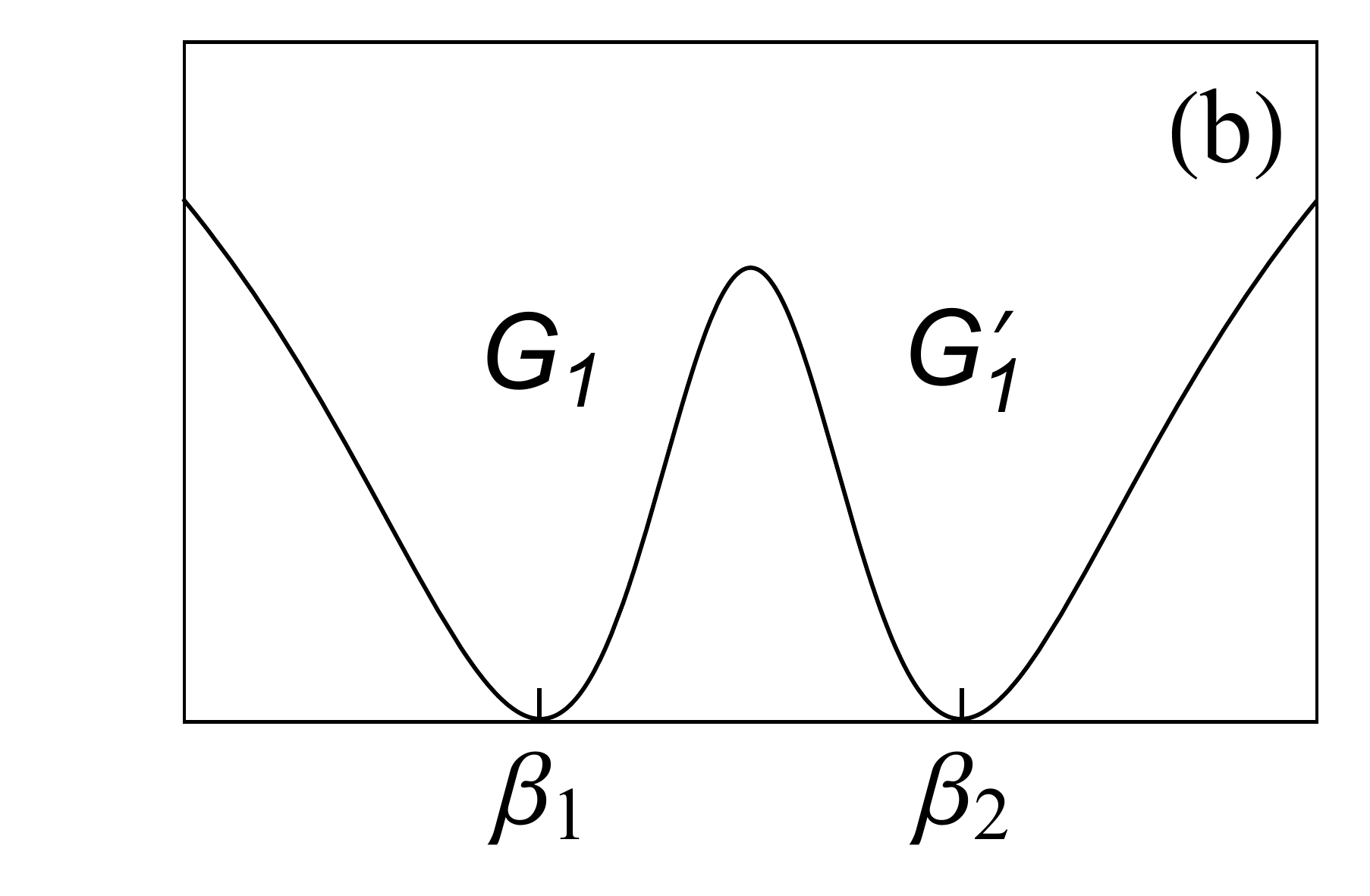}
\hspace{-0.4cm}
\includegraphics[width=5.6cm]{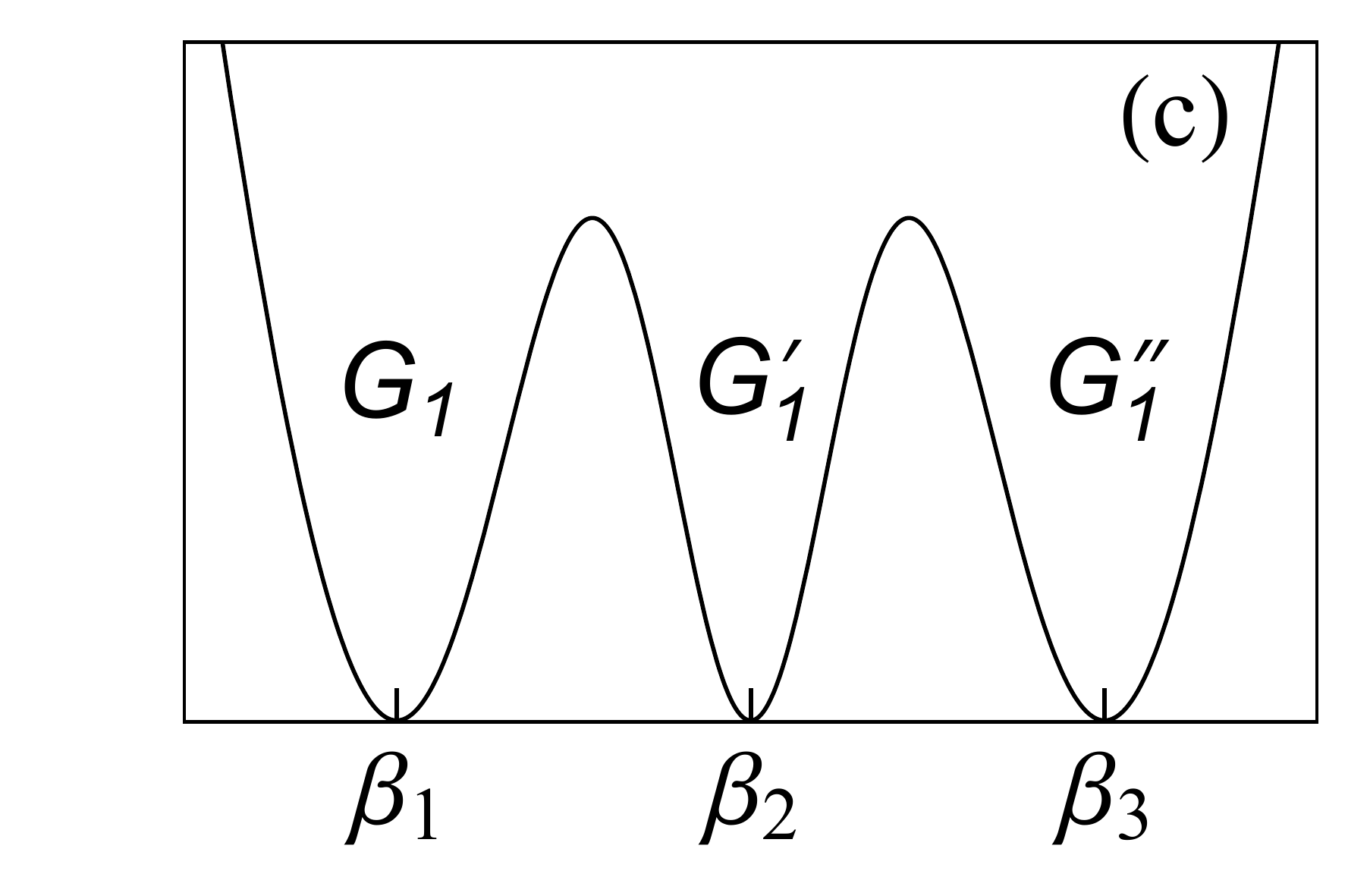}
\caption{\label{fig1}
\small
Energy surfaces accommodating: 
(a)~single minimum, (b)~double minima, 
(c)~triple minima, associated with 
(a)~$G_1$-DS or a single $G_1$-PDS, 
(b)~coexisting $G_1$-PDS and $G_1'$-PDS,
(c)~coexisting $G_1$-PDS, $G_1'$-PDS and $G_{1}''$-PDS.}
\label{fig1}
\end{figure}

A dynamical symmetry corresponds to a single structural phase with 
a particular shape $(\beta_{\rm eq},\gamma_{\rm eq})$. 
The DS Hamiltonians support a single minimum in their energy surface, 
hence serve as benchmarks for the dynamics of a single shape. 
Coexistence of different shapes in the same system is a ubiquitous phenomena 
in many-body systems, such as nuclei~\cite{Heyde11}. It involves the 
occurrence in the spectrum of several states (or bands of states) at similar 
energies with distinct properties, reflecting the nature of their 
dissimilar dynamics. 
The relevant Hamiltonians, by necessity, contain competing terms 
with incompatible (non-commuting) symmetries. The corresponding energy 
surface accommodates multiple minima, with different symmetry character 
(denoted by $G_1,\,G_{1}',\,G_{1}''$ in Fig.~1) of the dynamics 
in their vicinity. 
In such circumstances, exact DSs are broken, and any remaining 
symmetries can at most be shared by only a subset of states. 
To address the persisting regularities in such circumstances, amidst a 
complicated environment of other states, one needs to enlarge the 
traditional concept of exact dynamical symmetry. 
In the present contribution, we consider such an extended notion of 
symmetry, called partial dynamical symmetry (PDS)~\cite{Leviatan11} 
and show its potential role in formulating algebraic benchmarks for 
the dynamics of multiple shapes. 

\section{Partial dynamical symmetries} 
\label{PDSs}

A dynamical symmetry (DS) is characterized by {\it complete} solvability 
and good quantum numbers for {\it all} states. Partial dynamical 
symmetry (PDS)~\cite{Leviatan11,Alhassid92,Leviatan96} 
is a generalization of the latter concept, and corresponds 
to a particular symmetry breaking for which 
only {\it some} of the states retain solvability 
and/or have good quantum numbers. 
Such generalized forms of symmetries are manifested in nuclear structure, 
where extensive tests provide empirical evidence for their 
relevance to a broad range of 
nuclei~\cite{Leviatan11,Leviatan96,LevSin99,Casten14,Couture15,Casten16,
LevIsa02,Kremer14,LevGino00,GarciaRamos09,Leviatan13,Pds-BF15,
Isacker99,Escher00,Escher02,Rowe01,Rosen03,Isacker08,Isacker14}. 
In addition to nuclear spectroscopy, Hamiltonians with PDS 
have been used in the study of 
quantum phase transitions~\cite{Leviatan07,Macek14,LevDek16,LevGav17} 
and of systems with mixed regular and chaotic dynamics~\cite{WAL93,LW93}. 
In what follows, we present concrete algorithms for constructing 
Hamiltonians with single and multiple PDSs, and show 
explicit examples of Hamiltonians with such property, 
in the framework of the IBM.

The construction employs an intrinsic-collective resolution of the 
Hamiltonian~\cite{kirlev85,Leviatan87,levkir90,Leviatan06},
\ba
\hat{H}' &=& \hat{H} + \hat{H}_c ~.
\label{H-PDS}
\ea 
The intrinsic part ($\hat{H}$) determines the energy surface and 
band structure, while the collective part ($\hat{H}_c$) is composed of 
kinetic rotational terms and determines the in-band structure.
For a given shape, specified by the equilibrium deformations 
$(\beq,\gaeq)$, the intrinsic Hamiltonian is required 
to annihilate the equilibrium intrinsic state of Eq.~(\ref{int-state}),
\ba
\hat{H}\ket{\beq,\gaeq;N} = 0 ~.
\label{Ncond}
\ea
Since the Hamiltonian is rotational-invariant, 
this condition is equivalent to the requirement 
that $\hat{H}$ annihilates the states of good 
angular momentum~$L$ projected from $\ket{\beq,\gaeq;N}$
\ba
\hat{H}\ket{\beq,\gaeq;N,x,L} = 0 ~.
\label{Nxl}
\ea
Here $x$ denotes additional quantum numbers needed to characterize the 
states. Symmetry considerations enter when $(\beq,\gaeq)$ coincide 
with the equilibrium deformations of the DS chains listed in Table~1.
In this case, $\ket{\beq,\gaeq;N}$ 
becomes an extremal state in a particular irrep of the leading 
sub-algebra in the DS-chain and, consequently, the states 
projected from it have good symmetry. 

\subsection{Construction of Hamiltonians with a single PDS}

The starting point in constructing an Hamiltonian with 
a single PDS, is a dynamical symmetry chain,
\ba
{\rm U(6)\supset G_1\supset G_2\supset \ldots \supset SO(3)} &&\;\;\quad
\ket{N,\, \lambda_1,\,\lambda_2,\,\ldots,\,L} 
\;\;\qquad (\beq,\gaeq) ~,\quad 
\label{u6-ds}
\ea  
with leading sub-algebra $G_1$, related basis 
$\ket{N,\, \lambda_1,\,\lambda_2,\,\ldots,\,L}$ and associated shape 
$(\beta_{\rm eq},\gamma_{\rm eq})$ [see Eq.~(\ref{IBMchains}) and Table~1].
The PDS Hamiltonian of Eq.~(\ref{H-PDS}) is obtained in a two-step process. 
First, one identifies an intrinsic Hamiltonian ($\hat{H}$) 
with partial $G_1$-symmetry and then one identifies a 
collective Hamiltonian ($\hat{H}_c$), which ensures that the 
complete Hamiltonian ($\hat{H}'$) has $G_1$-PDS. 
For that purpose, the intrinsic Hamiltonian is required to satisfy 
\ba
\hat{H}\ket{\beq,\gaeq;N,\lambda_1\!=\!\Lambda_0,\lambda_2,\ldots,L} = 0 ~.
\label{Hvanish}
\ea 
The set of zero-energy eigenstates in Eq.~(\ref{Hvanish}) 
are basis states of a particular $G_1$-irrep, $\lambda_1=\Lambda_0$, 
with good $G_1$ symmetry, and are specified by the quantum numbers 
of the algebras in the chain~(\ref{u6-ds}). 
For a positive-definite $\hat{H}$, they span the ground band of 
the equilibrium shape and can be obtained by $L$-projection from 
the corresponding intrinsic state, 
$\vert\beta_{\rm eq},\gamma_{\rm eq} ; N\rangle$ of Eq.~(\ref{int-state}).
$\hat{H}$ itself, however, need not be invariant 
under $G_1$ and, therefore, has partial-$G_1$ symmetry. 
According to the PDS algorithms~\cite{Alhassid92,GarciaRamos09}, 
the construction of number-conserving Hamiltonians obeying the condition  
of Eq~(\ref{Hvanish}), is facilitated by writing them in 
normal-order form, 
\ba
\hat{H} &=& 
\sum_{\alpha,\beta}u_{\alpha\beta}\hat{T}^{\dag}_{\alpha}\hat{T}_{\beta} ~,
\label{H-normal}
\ea
in terms of $n$-particle creation and annihilation operators satisfying 
\ba
\hat{T}_{\alpha}
\ket{\beq,\gaeq;N,\lambda_1\!=\!\Lambda_0,\lambda_2,\ldots,L} = 0 ~.
\label{Talpha}
\ea
The collective part ($\hat{H}_c$) is identified with~the 
Casimir operators of 
the remaining sub-algebras of $G_1$ in the chain~(\ref{u6-ds}),
\ba
\hat{H}_c &=& \sum_{G_i\subset G_1} a_{G_i}\hat{C}[G_i] ~.
\label{H-col}
\ea
For this choice, the degeneracy of the above set of states 
is lifted, and they remain solvable eigenstates of the complete Hamiltonian 
$\hat{H}'$~(\ref{H-PDS}). The latter, by definition, has $G_1$-PDS.
It is interesting to note that $\hat{H}'=\hat{H}+\hat{H}_c$ has a form 
similar to the DS Hamiltonian, Eq.~(\ref{H-DS}), with the intrinsic 
part $\hat{H}$ replacing the Casimir operator of the leading sub-algebra, 
$\hat{C}[G_1]$ and $\hat{H}_c$ is diagonal and leads to splitting but no 
mixing. The difference, however, is that $\hat{C}[G_1]$ has all 
members of the DS basis $\ket{N,\, \lambda_1,\,\lambda_2,\,\ldots,\,L}$ 
as eigenstates, while $\hat{H}$ has only a subset of these basis states 
as eigenstates.

To demonstrate the above procedure, consider the SU(3)-DS chain of 
Eq.~(\ref{SU3-ds}). The two-boson pair operators 
\bsub
\ba
P^{\dagger}_{0} &=& d^{\dagger}\cdot d^{\dagger} - 2(s^{\dagger})^2 ~,
\label{P0}
\\
P^{\dagger}_{2m} &=& 2d^{\dagger}_{m}s^{\dagger} + 
\sqrt{7}\, (d^{\dagger}\,d^{\dagger})^{(2)}_{m} ~,
\label{P2}
\ea
\esub
are $(\lambda,\mu)=(0,2)$ tensors of SU(3) 
and annihilate the states of the SU(3) ground band,
\ba
P_{0}\,\ket{N,\, (\lambda,\mu)\!=\!(2N,0),\,K\!=\!0,\, L} &=& 0 ~,
\qquad\quad L=0,2,4\ldots,2N
\nonumber\\
P_{2m}\,\ket{N,\, (\lambda,\mu)\!=\!(2N,0),\,K\!=\!0,\, L} &=& 0 ~.
\label{P0P2}
\ea 
The operators of Eq.~(\ref{P0P2}) correspond to 
$\hat{T}_{\alpha}$ of Eq.~(\ref{Talpha}). The intrinsic Hamiltonian reads
\ba
\hat{H} = h_{0}\,P^{\dagger}_{0} P_0 
+ h_{2}\,P^{\dagger}_{2}\cdot \tilde{P}_{2} ~, 
\label{H-SU3PDS}
\ea
where $\tilde{P}_{2m} = (-)^{m}P_{2,-m}$ 
and the centered dot denotes a scalar product.
$\hat{H}$ of Eq.~(\ref{H-SU3PDS}) has partial SU(3) symmetry. 
For $h_2\!=\!h_0$ it is related to the quadratic Casimir operator of SU(3),
\ba
\hat{\theta}_2 \equiv
P^{\dagger}_{0}P_{0} + P^{\dagger}_{2}\cdot \tilde{P}_{2}  = 
-\hat C_{2}[{\rm SU(3)}] + 2\hat{N} (2\hat{N} +3) ~.
\label{theta2}
\ea
The collective Hamiltonian, $\hat{H}_c$, is composed of 
$\hat{C}_{2}[{\rm SO(3)}]$. The complete Hamiltonian 
$\hat{H}'$~(\ref{H-PDS}) has SU(3)-PDS, shown to be 
relevant to the spectroscopy of rare earth and actinide 
nuclei~\cite{Leviatan96,LevSin99,Casten14,Couture15,Casten16}.

A similar procedure can be adapted to the $\bsu3$-DS chain of 
Eq.~(\ref{SU3bar-ds}). The two-boson pair operators $P^{\dag}_0$ of 
Eq.~(\ref{P0}) and  
\ba
\bar{P}^{\dagger}_{2m} &=& -2d^{\dagger}_{m}s^{\dagger} + 
\sqrt{7}\, (d^{\dagger}\,d^{\dagger})^{(2)}_{m} ~,
\label{Pbar2}
\ea
are $(\blam,\bmu)=(2,0)$ tensors of $\bsu3$ and 
annihilate the states of the $\bsu3$ ground band,
\ba
P_{0}\,\ket{N,\, (\blam,\bmu)\!=\!(0,2N),\,\bar{K}\!=\!0,\, L} &=& 0 ~,
\qquad\quad L=0,2,4\ldots,2N
\nonumber\\
\bar{P}_{2m}\,\ket{N,\, (\blam,\bmu)\!=\!(0,2N),\,\bar{K}\!=\!0,\, L} &=& 0 ~.
\label{P0Pbar2}
\ea 
The intrinsic Hamiltonian reads
\ba
\hat{H} = t_{0}\,P^{\dagger}_{0} P_0 
+ t_{2}\,\bar{P}^{\dagger}_{2}\cdot \tilde{\bar{P}}_{2} ~, 
\label{H-SU3bPDS}
\ea
and has partial $\bsu3$ symmetry. For $t_2=t_0$ it 
is related to the Casimir operator of $\bsu3$,
\ba
\hat{\bar{\theta}}_2 \equiv
P^{\dagger}_{0}P_{0} + \bar{P}^{\dagger}_{2}\cdot \tilde{\bar{P}}_{2}  = 
-\hat C_{2}[{\rm \bsu3}] + 2\hat{N} (2\hat{N} +3) ~.
\label{thetab2}
\ea

Occasionally, the condition of Eq.~(\ref{Hvanish}) necessitates higher-order 
terms in the Hamiltonian. This is the case for the SO(6)-DS 
of Eq.~(\ref{SO6-ds}). 
Here the two-boson pair operator 
\ba
R^{\dagger}_{0} &=& d^{\dagger}\cdot d^{\dagger} - (s^{\dagger})^2 ~,
\label{R0}
\ea
is a scalar ($\sigma=0$) under SO(6) and 
annihilates the states in the SO(6) ground-band,
\ba
R_{0}\,\vert N,\sigma=N,\tau, n_{\Delta}, L\rangle &=& 0 ~,
\qquad\quad \tau=0,1,2,\ldots,N
\label{R0vanish}
\ea
However, in this case, the following two-body term 
\ba
\hat{\theta}_0 \equiv R^{\dag}_0R_0 =
-\hat C_{2}[{\rm SO(6)}] + \hat{N} (\hat{N}+4) ~.
\label{theta0}
\ea
is related to the Casimir operator of SO(6), hence is SO(6)-invariant. 
A genuine SO(6)-PDS  can be realized by including cubic terms in 
the intrinsic Hamiltonian,
\ba
\hat{H} = 
r_0\,R^{\dag}_0\hat{n}_sR_0 + r_2\,R^{\dag}_0\hat{n}_dR_0 ~.
\label{HintSO6}
\ea
The collective Hamiltonian, $\hat{H}_c$, is composed of 
$\hat{C}_{2}[{\rm SO(5)}]$ and $\hat{C}_{2}[{\rm SO(3)}]$. 
The complete Hamiltonian 
$\hat{H}'$ (\ref{H-PDS}) has SO(6)-PDS, shown to be 
relevant to the spectroscopy of $^{196}$Pt~\cite{GarciaRamos09}.

\subsection{Construction of Hamiltonians with several PDSs}
\label{singlePDS}

The procedure described in the previous section was oriented towards 
constructing Hamiltonians with a single PDS. 
A large number of such PDS 
Hamiltonians~\cite{Leviatan96,LevSin99,Casten14,Couture15,Casten16,
LevIsa02,Kremer14,LevGino00,GarciaRamos09,Leviatan13,Pds-BF15} 
have been constructed in this manner. They provide a valuable 
addition to the arsenal of DS Hamiltonians, suitable for describing
systems with a single shape. To allow for a description of 
multiple shapes in the same system, requires an extension of 
the above procedure to encompass a construction of Hamiltonians with 
several distinct PDSs~\cite{Leviatan07,Macek14,LevDek16,LevGav17}. 
This is the subject matter of the present contribution. 
We focus on the dynamics in the vicinity of the critical 
point, where the corresponding multiple minima in the energy surface 
are near-degenerate and the structure changes most rapidly.

For that purpose, consider two different shapes 
specified by equilibrium deformations 
($\beta_1,\gamma_1$) and ($\beta_2,\gamma_2$) 
whose dynamics is described, respectively, by the following DS chains
\bsub
\ba
{\rm U(6)\supset G_1\supset G_2\supset \ldots \supset SO(3)} &&\;\;\quad
\ket{N,\, \lambda_1,\,\lambda_2,\,\ldots,\,L} 
\;\;\qquad (\beta_1,\gamma_1) ~,\quad 
\label{ds-G1}\\
{\rm U(6)\supset G'_1\supset G'_2\supset \ldots \supset SO(3)} &&\;\;\quad
\ket{N,\, \sigma_1,\,\sigma_2,\,\ldots,\,L} 
\;\;\qquad (\beta_2,\gamma_2) ~,\quad 
\label{ds-G1prime}
\ea
\esub
with different leading sub-algebras ($G_1\neq G'_1$) and associated bases.
As portrayed in Fig.~1, at the critical point, the corresponding minima 
representing the two shapes, and respective ground bands are degenerate. 
Accordingly, we require the intrinsic critical-point Hamiltonian to satisfy 
simultaneously the following two conditions
\bsub
\ba
\hat{H}\ket{\beta_1,\gamma_1;N,\lambda_1\!=\Lambda_0,\lambda_2,\ldots,L} 
&=& 0 ~,
\label{basis1}\\
\hat{H}\ket{\beta_2,\gamma_2;N,\sigma_1=\Sigma_0,\sigma_2,\ldots,L} 
&=&0 ~.
\label{basis2}
\ea
\label{bases12}
\esub
The states of Eq.~(\ref{basis1}) reside in the $\lambda_1=\Lambda_0$ irrep 
of $G_1$, are classified according to the DS-chain (\ref{ds-G1}), hence 
have good $G_1$ symmetry. Similarly,  
the states of Eq.~(\ref{basis2}) reside in the $\sigma_1=\Sigma_0$ irrep 
of $G'_1$, are classified according to the DS-chain (\ref{ds-G1prime}), 
hence have good $G'_1$ symmetry. Although $G_1$ and $G'_1$ are incompatible, 
both sets are eigenstates of the same Hamiltonian. When the latter 
is positive definite, the two sets span the ground bands of the 
$(\beta_1,\gamma_1)$ and $(\beta_2,\gamma_2)$ shapes, respectively.
In general, $\hat{H}$ itself is not necessarily 
invariant under $G_1$ nor under $G_2$ and, therefore, its other eigenstates 
can be mixed with respect to both $G_1$ and $G'_1$. 
Identifying the collective part of the Hamiltonian with the Casimir 
operator of SO(3) (as well as with the Casimir operators of additional 
algebras which are common to both chains), the two sets of states 
remain (non-degenerate) eigenstates of the complete 
Hamiltonian~(\ref{H-PDS}), which 
then has both $G_1$-PDS and $G'_1$-PDS. 
The case of triple (or multiple) 
shape coexistence, associated with three (or more) incompatible DS-chains is 
treated in a similar fashion. 
In the following sections, we apply the above procedure 
to a variety of coexisting shapes in the IBM framework, 
examine the spectral properties 
of the derived PDS Hamiltonians, and highlight their potential to serve 
as benchmarks for describing multiple shapes in nuclei. 

\section{Simultaneous occurrence of a spherical shape and a deformed shape}
\label{spher-def}

A particular type of shape coexistence present in nuclei 
involves spherical and quadrupole-deformed shapes,
{\it e.g.}, in neutron-rich Sr isotopes~\cite{Clement16,Park16}, 
$^{96}$Zr~\cite{kremer16} and near $^{78}$Ni~\cite{gottardo16,yang16}). 
A PDS Hamiltonian appropriate 
to simultaneously occurring spherical and axially-deformed prolate shapes, 
can be obtained from Eq.~(\ref{H-SU3PDS}) by setting $h_0\!=\!0$. 
This Hamiltonian was studied in great detail in~\cite{Leviatan07,Macek14}, 
and shown to have coexisting U(5)-PDS and SU(3)-PDS. In a similar manner, 
a PDS Hamiltonian with coexisting U(5)-PDS and $\bsu3$-PDS, appropriate 
to simultaneously occurring spherical and axially-deformed oblate shapes, 
can be obtained from Eq.~(\ref{H-SU3bPDS}) by setting $t_0\!=\!0$. 
The $\gamma$ degree of freedom and triaxiality can play an 
important role in the occurrence of multiple shapes in nuclei~\cite{Ayan16}.
In what follows, we consider a case study of PDS Hamiltonians 
relevant to a coexistence of spherical and non-axial $\gamma$-unstable 
deformed shapes.

 \subsection{Coexisting U(5)-PDS and SO(6)-PDS} 
\label{SGshapes}

The U(5)-DS limit of Eq.~(\ref{U5-ds}) 
is appropriate to the dynamics of a spherical shape. 
For a given $N$, the allowed U(5) and SO(5) irreps 
are $n_d\!=\!0,1,2,\ldots, N$ and  $\tau\!=\!n_d,\,n_d\!-\!2,\dots 0$ 
or~$1$, respectively. The values of $L$ contained in a given $\tau$-irrep 
follow the ${\rm SO(5)\supset SO(3)}$ reduction rules~\cite{ibm}. 
The U(5)-DS spectrum resembles that of an anharmonic spherical vibrator, 
composed of U(5) $n_d$-multiplets whose spacing is governed by 
$\hat{C}_{1}[{\rm U(5)}]\!=\!\hat{n}_d$, 
and the splitting is generated by the SO(5) and SO(3) terms. 
The lowest U(5) multiplets involve the ground state 
with quantum numbers $(n_d\!=\!0,\,\tau\!=\!0,\, L\!=\!0)$ 
and excited states with quantum numbers 
$(n_d=\!1\!,\,\tau\!=\!1,\, L\!=\!2)$, 
$(n_d\!=\!2:\,\tau\!=\!0,\,L\!=\!0;\,\tau\!=\!2,\,L\!=\!2,4)$ and 
$(n_d\!=\!3:\,\tau\!=\!3,\,L\!=\!6,4,3,0;\,\tau\!=\!1,\,L\!=\!2)$.

The SO(6)-DS limit of Eq.~(\ref{SO6-ds}) is appropriate to the 
dynamics of a $\gamma$-unstable deformed shape. 
For a given $N$, the allowed SO(6) and SO(5) irreps are 
$\sigma\!=\!N,\,N-2,\dots 0$ or $1$, and  
$\tau=0,\,1,\,\ldots \sigma$, respectively. 
The ${\rm SO(5)}\supset {\rm SO(3)}$ reduction is the same as in the 
U(5)-DS chain. 
The SO(6)-DS spectrum resembles that of a $\gamma$-unstable deformed 
rotovibrator, composed of SO(6) $\sigma$-multiplets forming rotational 
bands, with $\tau(\tau+3)$ and $L(L+1)$ splitting generated 
by $\hat{C}_{2}[{\rm SO(5)}]$ and 
$\hat{C}_{2}[{\rm SO(3)}]$, respectively. 
The lowest irrep $\sigma\!=\!N$ contains the ground ($g$) band  
of a $\gamma$-unstable deformed nucleus. 
The first excited irrep $\sigma\!=\!N-2$ contains the $\beta$-band. 
The lowest members in each band have quantum numbers 
$(\tau=0,\, L=0)$, $(\tau=1,\, L=2)$, 
$(\tau=2,\, L=2,4)$ and $(\tau=3,\, L=0,3,4,6)$. 

Following the procedure outlined in Eq.~(\ref{bases12}),
the intrinsic part of the 
critical-point Hamiltonian, relevant to spherical and $\gamma$-unstable 
deformed (S-G) shape-coexistence, is required to satisfy
\bsub
\ba
\hat{H}\ket{N,\, \sigma=N,\,\tau,\, L} &=& 0 ~,
\label{sigmaN}
\\
\hat{H}\ket{N,\, n_d=0,\,\tau=0,\,L=0} &=& 0 ~.
\label{nd0SG}
\ea
\label{o6u5solv}
\esub 
Equivalently, $\hat{H}$ annihilates both the deformed intrinsic state 
of Eq.~(\ref{int-state}) with $(\beta=1,\gamma\,{\rm arbitrary})$,
which is the lowest weight vector in the SO(6) irrep
$\sigma\!=\!N$, and the spherical intrinsic state with $\beta=0$, 
which is the single basis state in the U(5) irrep $n_d\!=\!0$. 
The resulting intrinsic Hamiltonian is found to be that 
of Eq.~(\ref{HintSO6}) with $r_0=0$, 
\ba
\hat{H} = 
r_2\,R^{\dag}_0\hat{n}_dR_0 ~,
\label{HintSG}
\ea
and $R^{\dag}_0$ given in Eq.~(\ref{R0}). 
\begin{figure}[t]
\centering
\includegraphics[width=16cm]{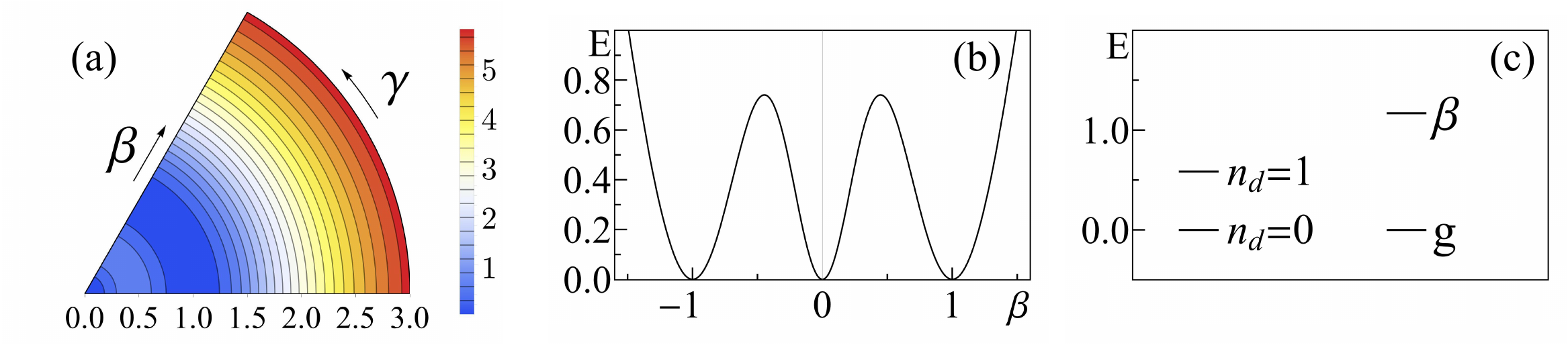}
\caption{\label{fig2}
\small
Spherical and $\gamma$-unstable deformed (S-G) shape coexistence.
(a)~Contour plots of the $\gamma$-independent 
energy surface~(\ref{surfaceSG}),   
(b)~$\gamma\!=\!0$ sections, and
(c)~bandhead spectrum, for the Hamiltonian $\hat{H}'$~(\ref{HprimeSG})
with parameters $r_2\!=\!10,\,\rho_5\!=\!\rho_3=0$ and $N\!=\!20$.}
\end{figure}
The energy surface, 
$E_{N}(\beta,\gamma) = N(N-1)(N-2)\tilde{E}(\beta,\gamma)$, 
is given by
\ba
\tilde{E}(\beta) = 
r_2\beta^2(\beta^2-1)^2 (1+\beta^2)^{-3} ~.
\label{surfaceSG}
\ea
The surface is an even sextic function of $\beta$ and is independent 
of $\gamma$, in accord with the SO(5) symmetry of the Hamiltonian. 
For $r_2 > 0$, 
$\hat{H}$ is positive definite and 
$\tilde{E}(\beta)$ has two degenerate global minima, 
$\beta\!=\!0$ and $\beta^2\!=\!1$, at $\tilde{E}\!=\!0$.
A local maximum at $\bs^2\!=\!\frac{1}{5}$ creates a barrier 
of height $\tilde{E} \!=\!\frac{2}{27}r_2$, separating the two minima, 
as seen in Fig.~2. 
For large $N$, the normal modes shown schematically in Fig.~2(c), 
involve $\beta$ vibrations about the deformed minima, with frequency 
$\epsilon_{\beta}$, and quadrupole vibrations about the spherical minimum, 
with frequency $\epsilon$, respectively, 
\bsub
\ba
\epsilon_{\beta} &=& 2r_2\,N^2 ~,
\\
\epsilon &=& r_2\,N^2 ~.
\label{sg-modes}
\ea
\esub
\begin{figure}[t]
  \centering
\begin{minipage}{17pc}
\vspace{-0.1cm}
\includegraphics[width=7.6cm]{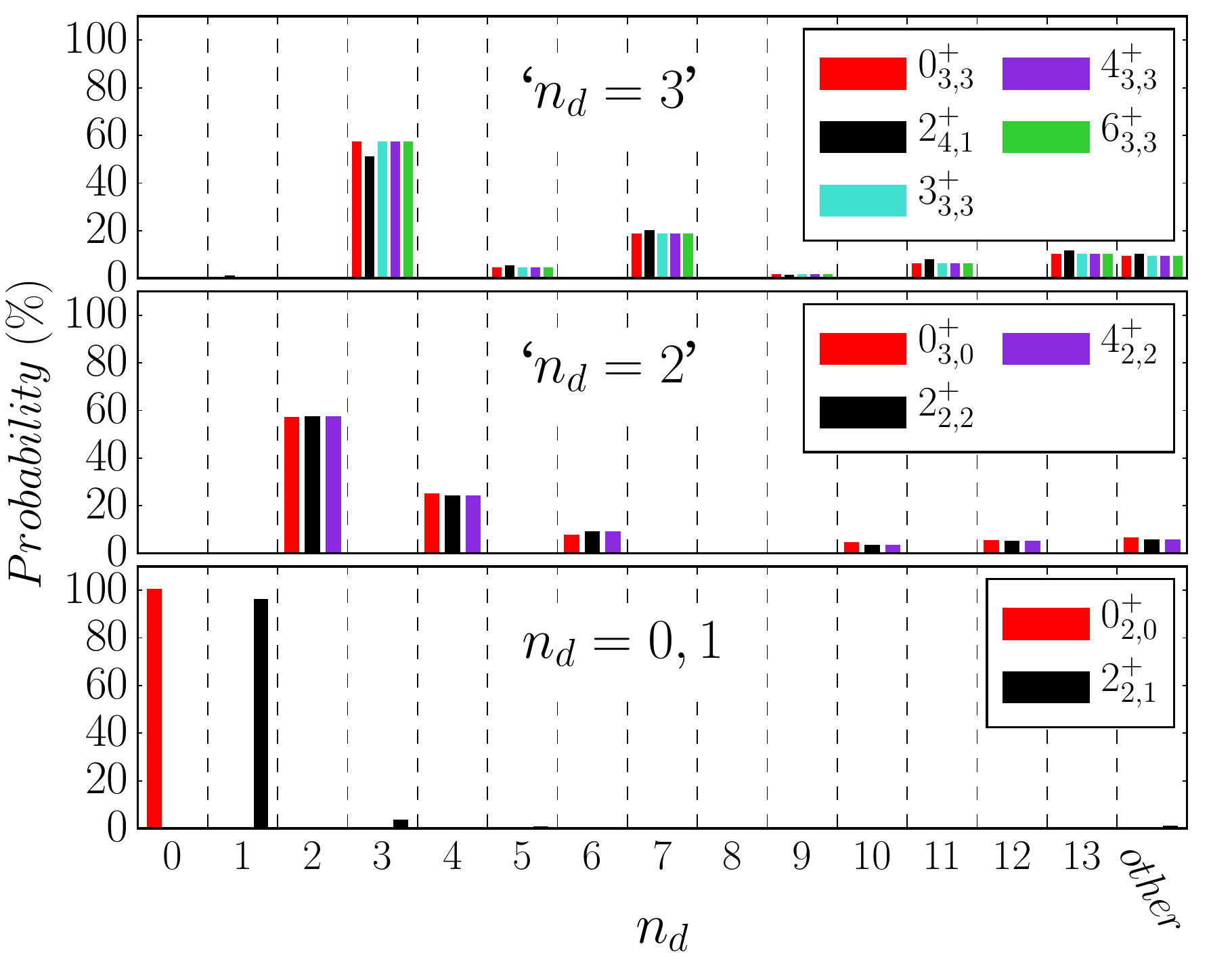}
\caption{\label{fig3-SG}
\small
U(5) $n_d$-decomposition for spherical states, 
eigenstates of 
the Hamiltonian $\hat{H}'$~(\ref{HprimeSG}) 
with parameters as in Fig.~2, resulting in 
spherical and $\gamma$-unstable deformed (S-G) shape coexistence.
The column `other' depicts a sum of probabilities, each less than~5\%.
The spherical states are dominated by a single $n_d$ component, in marked 
contrast to the deformed states exhibiting a broad $n_d$-distribution.} 
%(see Fig.~4).}
%thus signaling the presence in the spectrum of U(5)-PDS.} 
\end{minipage}\hspace{0.7cm}%
\begin{minipage}{18pc}
\includegraphics[width=7.6cm]{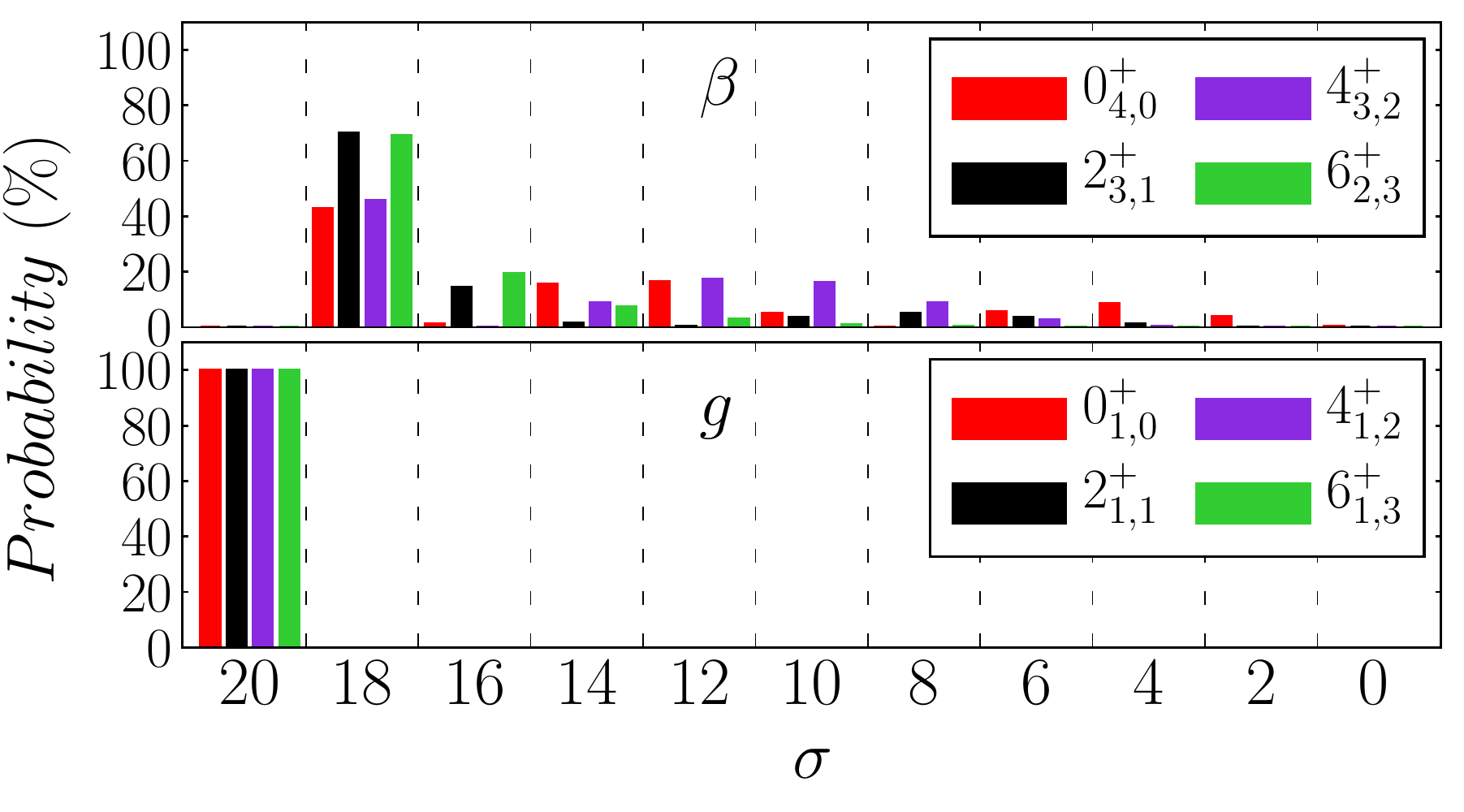}
\includegraphics[width=7.6cm]{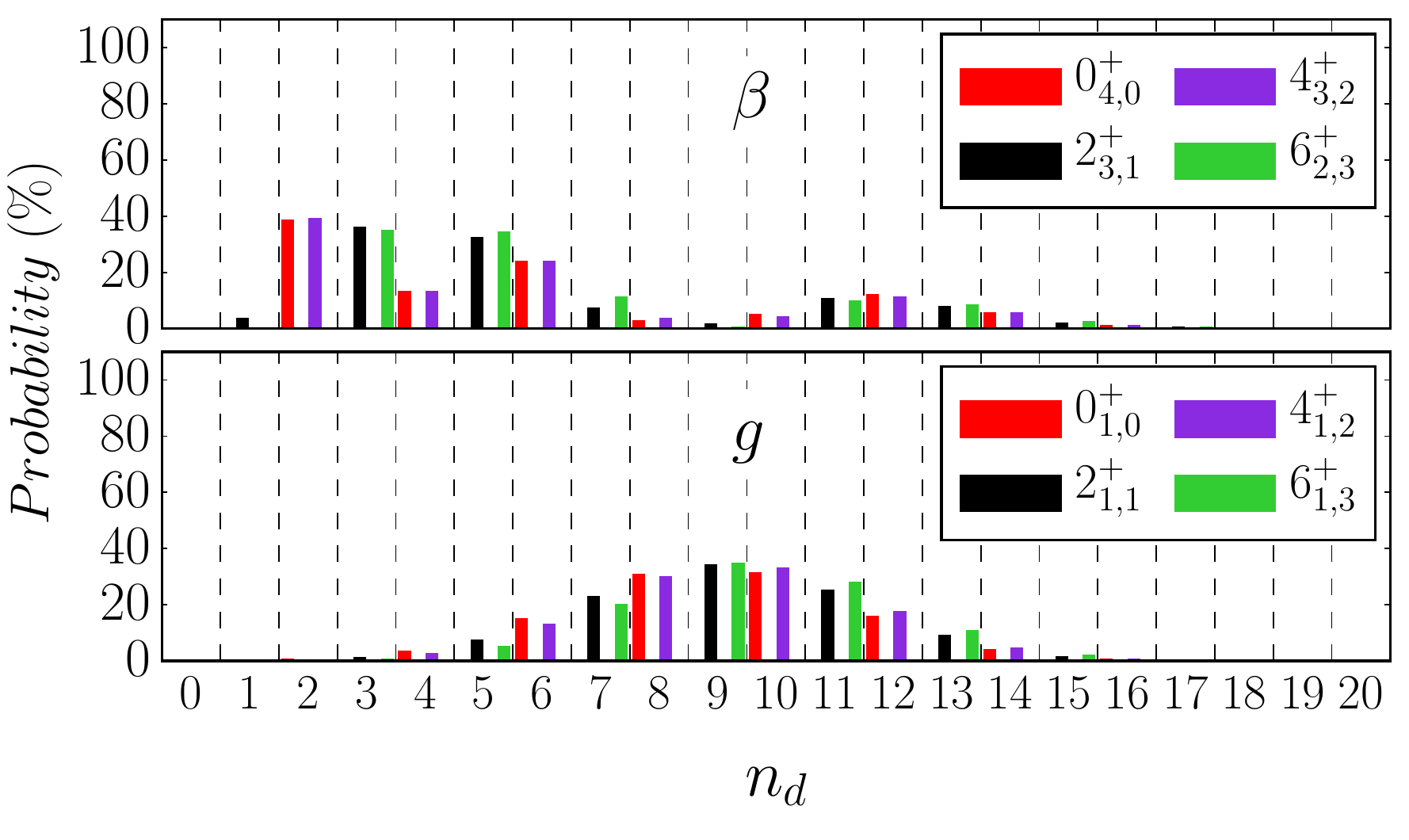}
\caption{\label{fig4}
\small
SO(6) $\sigma$- and U(5) $n_d$- decomposition 
for members of the deformed ground ($g$) and $\beta$ bands, 
eigenstates of $\hat{H}'$~(\ref{HprimeSG}) as in Fig.~2.}
\end{minipage} 
\end{figure} 

\vspace{-12pt}
Identifying the collective part of the 
Hamiltonian with the 
Casimir operators of the common ${\rm SO(5)\supset SO(3)}$ segment 
of the chains~(\ref{U5-ds}) and (\ref{SO6-ds}), 
we arrive at the following complete Hamiltonian 
\ba
\hat{H}' &=& r_2\,R^{\dag}_0\hat{n}_dR_0
+ \rho_5\,\hat{C}_2[\rm SO(5)] + \rho_3\,\hat{C}_2[\rm SO(3)] ~.
\label{HprimeSG}
\ea 
The added rotational terms generate an exact 
$\rho_5\tau(\tau+3)+\rho_3L(L+1)$ splitting without 
affecting the wave functions.
In particular, the solvable subset of 
eigenstates, Eq.~(\ref{o6u5solv}), remain intact.
Since both SO(5) and SO(3) are preserved by the Hamiltonian, 
its eigenstates have good $(\tau,L)$ quantum numbers and can be labeled 
as $L^{+}_{i,\tau}$, where the ordinal number $i$ enumerates the occurrences 
of states with the same $(\tau,L)$, with increasing energy.
The nature of the Hamiltonian eigenstates can be inferred from the 
probability distributions, ${\textstyle P^{(N,\tau,L)}_{n_d} = 
\vert C_{n_d}^{(N,\tau,L)}\vert^2}$ and 
${\textstyle P^{(N,\tau,L)}_{\sigma} = 
\vert C_{\sigma}^{(N,\tau,L)}\vert^2}$, obtained from their expansion 
coefficients in the U(5) and SO(6) bases, Eqs~(\ref{U5-ds}) 
and (\ref{SO6-ds}). 
In general, the low lying spectrum of $\hat{H}'$~(\ref{HprimeSG}) 
exhibits two distinct classes of states. The first class consists 
of ($\tau,L$) states arranged in $n_d$-multiplets of a spherical vibrator.
Fig.~3 shows the U(5) $n_d$-decomposition 
of such spherical states, characterized by a narrow $n_d$-distribution. 
The lowest spherical state, $L=0^{+}_{2,0}$, is the solvable 
U(5) state of Eq.~(\ref{nd0SG}) with U(5) quantum number $n_d=0$. 
The $L=2^{+}_{2,1}$ state has $n_d=1$ to a good approximation.
The upper panels of Fig.~3 display the next spherical-type of 
multiplets ($L=0^{+}_{3,0},\,2^{+}_{2,2},\,4^{+}_{2,2}$)
and ($L=6^{+}_{3,3},\,4^{+}_{3,3},\,3^{+}_{3,3},\,0^{+}_{3,3},\,2^{+}_{4,1}$), 
which have a somewhat less pronounced (60\%)
single $n_d$-component, with $n_d=2$ and $n_d=3$, respectively.

A second class consists of ($\tau,L$) states arranged in 
bands of a $\gamma$-unstable deformed rotor. 
The SO(6) $\sigma$-decomposition of such states, in selected bands, 
are shown in the upper panel of Fig.~4.
The ground band is seen to be pure with $\sigma\!=\!N$ SO(6) character, 
and coincides with the solvable band of Eq.~(\ref{sigmaN}).
In contrast, the non-solvable $\beta$-band (and higher $\beta^n$-bands) 
show considerable SO(6)-mixing. The deformed nature of these SO(5)-rotational 
states is manifested in their broad $n_d$-distribution, shown in the 
lower panel of Fig.~4.
The above analysis demonstrates that although 
the critical-point Hamiltonian~(\ref{HprimeSG}) 
is not invariant under U(5) nor SO(6), 
some of its eigenstates have good U(5) symmetry, 
some have good SO(6) symmetry 
and all other states are mixed with respect to both U(5) and SO(6). 
These are precisely the defining attributes of U(5)-PDS coexisting 
with SO(6)-PDS.

Since the wave functions for the solvable states, 
Eqs.~(\ref{o6u5solv}), are known, one has at hand closed form 
expressions for related spectroscopic observables. 
Consider the $E2$ operator, 
\ba
T(E2) = e_B\,\Pi^{(2)} = e_B\,(d^{\dag}s + s^{\dag}\tilde{d}) ~,
\label{TE2}
\ea
where $\Pi^{(2)}$ is a generator of SO(6) 
and $e_B$ an effective charge. $T(E2)$ obeys the SO(5) selection rules 
$\Delta\tau=\pm 1$ and, consequently, all $(\tau,L)$ states have 
vanishing quadrupole moments. 
The $B(E2)$ values for intraband ($g\to g$) 
transitions between states of the ground band, Eq.~(\ref{sigmaN}), 
are given by the known SO(6)-DS expressions~\cite{ibm}. For example,
\bsub
\ba
B(E2;\, g,\, \tau+1,\,L'=2\tau+2\to g,\,\tau,\,L=2\tau) &=& 
{\textstyle
e_{B}^2\,\frac{\tau+1}{2\tau+5}
(N-\tau)(N+\tau+4)} ~,
\qquad
\label{be2So6a}\\[1mm]
B(E2;\, g,\, \tau+1,\,L'=2\tau\to g,\,\tau,\,L=2\tau) &=& 
{\textstyle
e_{B}^2\,\frac{4\tau+2}{(2\tau+5)(4\tau-1)}
(N-\tau)(N+\tau+4)} ~.
\qquad\quad
\label{be2So6b}
\ea
\label{be2So6}
\esub
Similarly, the $E2$ rates for the transition connecting the pure 
spherical states, $(n_d\!=\!\tau\!=\!1,L\!=\!2)$ and 
$(n_d\!=\!\tau\!=\!0,L\!=\!0)$, satisfy 
the U(5)-DS expression~\cite{ibm}
\ba
B(E2; n_d=1,L=2\to n_d=0,L=0) = e_{B}^2N ~.\quad
\label{be2nd}
\ea
Member states of the deformed ground band~(\ref{sigmaN}) 
span the entire $\sigma=N$ irrep of SO(6) and are not connected by 
$E2$ transitions to the spherical states since 
$\Pi^{(2)}$, as a generator of SO(6), cannot connect different 
$\sigma$-irreps of SO(6). The weak spherical $\to$ deformed 
$E2$ transitions persist also for a more general E2 operator 
obtained by adding $(d^{\dag}\tilde{d})^{(2)}$ to $T(E2)$, 
since the latter term, as a generator of U(5), cannot connect 
different $n_d$-irreps of U(5).
By similar arguments, there are 
no $E0$ transitions involving these spherical states, 
since the $E0$ operator, $T(E0)\propto \hat{n}_d$, is diagonal in $n_d$.
These symmetry-based selection rules result in strong electromagnetic 
transitions between states in the same class, associated with a given shape, 
and weak transitions between states in different classes. 
\begin{figure}[t]
\centering
\includegraphics[width=0.35\linewidth]{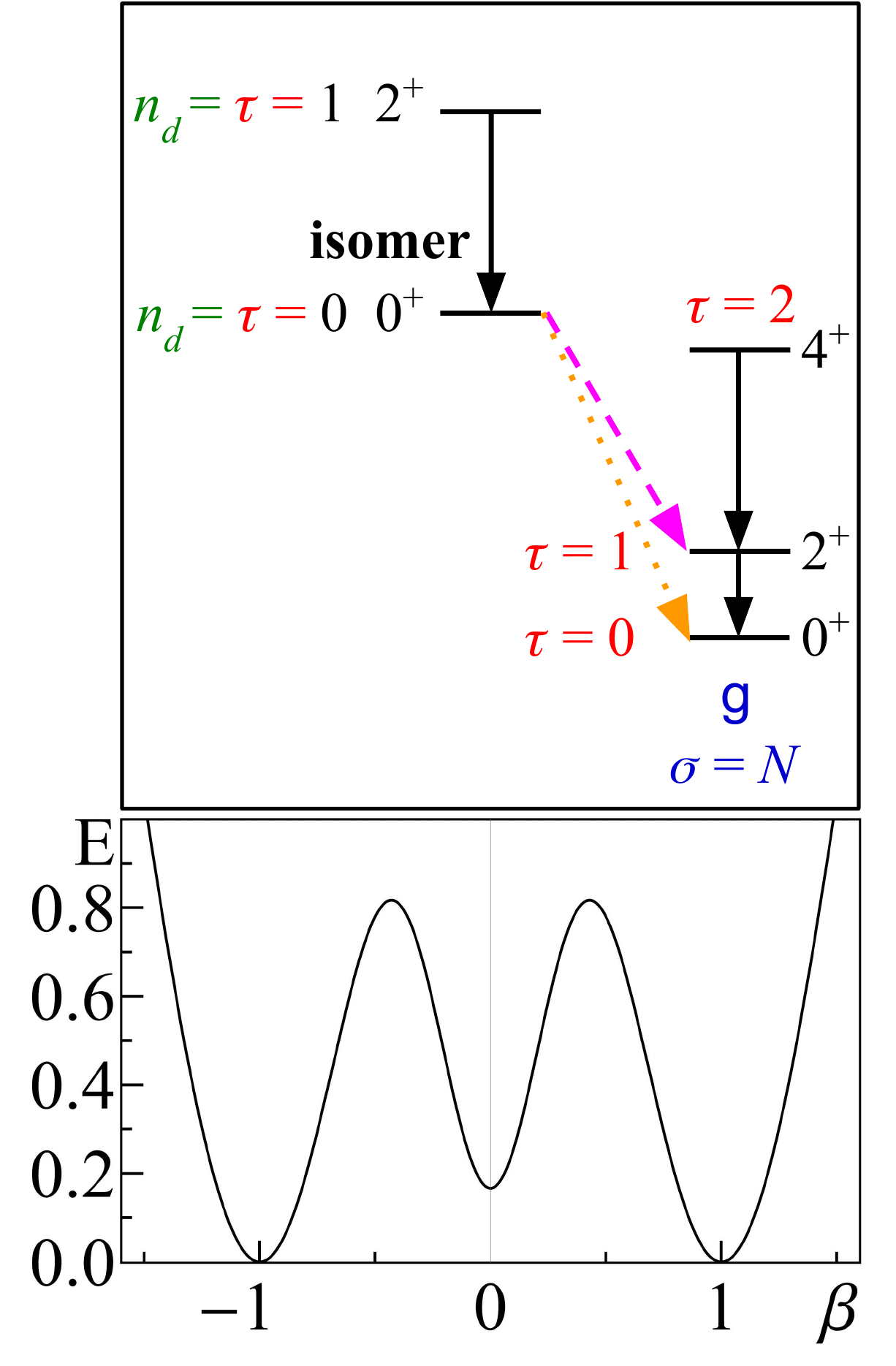}
\hspace{0.5cm}%
\includegraphics[width=0.35\linewidth]{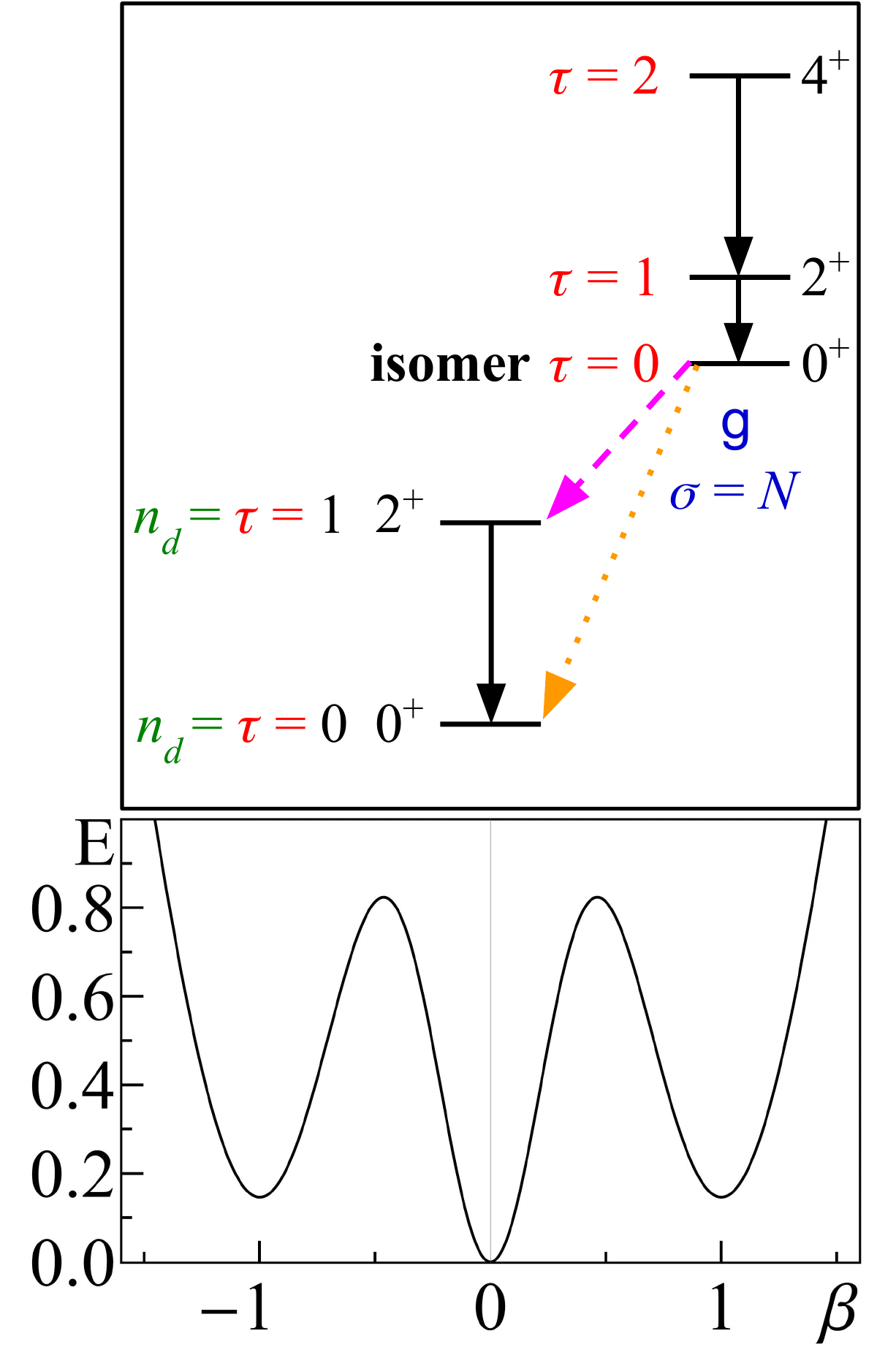}
\caption{\label{fig5-SG}
\small
Energy-surface sections and level schemes 
corresponding to departures from the critical point of S-G shape 
coexistence, for $\hat{H}'$~(\ref{HprimeSG}) with parameters as in Fig.~2.
Left panels: a spherical isomeric state 
($\alpha\hat{\theta}_0$ term~(\ref{theta0}), 
with $\alpha\!=\!3$, added to $\hat{H}'$).
Right panels: a $\gamma$-unstable deformed 
isomeric state ($\epsilon\hat{n}_d$ term, with $\epsilon\!=\!200$, 
added to $\hat{H}'$).
Retarded $E2$ (dashed lines) and $E0$ (dotted lines) decays identify 
the isomeric states.}
\end{figure}

The above discussion has focused on the dynamics in the vicinity of the 
critical point, 
where the spherical and deformed configurations are degenerate.
The evolution of structure away from the critical point, 
can be studied by adding to $\hat{H}'$~(\ref{HprimeSG}) 
the Casimir operators of U(5) and SO(6), 
still retaining the desired SO(5) symmetry. 
Adding an $\epsilon\hat{n}_d$ term, 
will leave the energy of the spherical $n_d=0$ state unchanged, 
but will shift the deformed $\gamma$-unstable ground band to 
higher energy of order $\epsilon N/2$. 
Similarly, adding a small $\alpha\hat{\theta}_0$ term (\ref{theta0}), 
will leave the solvable SO(6) $\sigma\!=\!N$ ground band unchanged, but will 
shift the spherical ground state ($n_d\!=\!L\!=\!0$) to higher energy 
of order $\alpha N^2$. 
The resulting topology of the energy surfaces with such modifications 
are shown at the bottom row of Fig.~5.
If these departures 
from the critical points are small, the wave functions decomposition of 
Figs.~3-4 remain intact and the above analytic 
expressions for $E2$ observables 
and selection rules are still valid to a good approximation.
In such scenarios, the lowest $L\!=\!0$ state of the non-yrast configuration 
will exhibit retarded $E2$ and $E0$ decays, hence will have the attributes 
of an isomer state, as depicted schematically on the top row 
of Fig.~5.

\section{Simultaneous occurrence of two deformed shapes}

Shape coexistence in nuclei can involve two deformed shapes 
({\it e.g.}, prolate and oblate) as encountered in Kr~\cite{Clement07}, 
Se~\cite{Ljun08} and Hg isotopes~\cite{Bree14}.
In what follows, we study PDS Hamiltonians relevant 
to such axially-deformed shapes with $\gamma=0,\pi/3$ and
equal $\beta$ deformation.

\subsection{Coexisting SU(3)-PDS and $\bsu3$-PDS} 
\label{POshapes}

The DS limits appropriate to prolate and oblate shapes correspond 
to the chains~(\ref{SU3-ds}) and (\ref{SU3bar-ds}), respectively. 
For a given $N$, the allowed SU(3) [$\,\bsu3\,$] irreps are 
$(\lambda,\mu)\!=\!(2N \!-\! 4k \!-\! 6m,2k)$ 
[$(\blam,\bmu)\!=\!(2k,2N\!-\!4k\!-\!6m)$] 
with $k,m$, non-negative integers. 
The multiplicity label $K$ ($\bK$) corresponds geometrically to the
projection of the angular momentum ($L$) on the symmetry axis. 
The basis states are eigenstates of the Casimir operator 
$\hat{C}_{2}[{\rm SU(3)}]$ or $\hat{C}_{2}[\bsu3]$, 
with eigenvalues listed in Table~1. 
Specifically, $\hat{C}_{2}[{\rm SU(3)}] \!=\! 2Q^{(2)}\cdot Q^{(2)} \!+\! 
{\textstyle\frac{3}{4}}L^{(1)}\cdot L^{(1)}$, 
$Q^{(2)} \!=\! d^{\dagger}s \!+\! s^{\dagger}\tilde{d} 
\!-\!\frac{1}{2}\sqrt{7} (d^{\dagger}\tilde{d})^{(2)}$, 
$L^{(1)} \!=\! \sqrt{10} (d^{\dagger}\tilde{d})^{(1)}$, 
and $\hat{C}_{2}[\bsu3]$ is obtained by replacing $Q^{(2)}$ by 
$\bar{Q}^{(2)} \!=\! d^{\dagger}s \!+\! s^{\dagger}\tilde{d} 
\!+\!\frac{1}{2}\sqrt{7} (d^{\dagger}\tilde{d})^{(2)}$. 
The generators of SU(3) and $\bsu3$, $Q^{(2)}$ and $\bar{Q}^{(2)}$, 
and corresponding basis states, are related 
by a change of phase $(s^{\dag},s)\rightarrow (-s^{\dag},-s)$, 
induced by the operator ${\cal R}_s=\exp(i\pi\hat{n}_s)$, 
with $\hat{n}_s=s^{\dag}s$. 
The DS spectrum resembles that of an axially-deformed 
rotovibrator composed of SU(3) [or $\bsu3$] multiplets forming 
rotational bands, with $L(L+1)$-splitting generated by 
$\hat{C}_{2}[{\rm SO(3)}] \!=\! L^{(1)}\cdot L^{(1)}$. 
In the SU(3) [or $\bsu3$] DS limit, the lowest irrep $(2N,0)$ [or $(0,2N)$] 
contains the ground band $g(K\!=\!0)$ [or $g(\bK\!=\!0)$] 
of a prolate [oblate] deformed nucleus. 
The first excited irrep $(2N\!-\!4,2)$ [or $(2,2N\!-\!4)$] contains 
both the $\beta(K\!=\!0)$ and $\gamma(K\!=\!2)$ 
[or $\beta(\bK\!=\!0)$ and $\gamma(\bK\!=\!2)$] bands. 
Henceforth, we denote such prolate and oblate bands by 
$(g_1,\beta_1,\gamma_1)$ and ($g_2,\beta_2,\gamma_2$), respectively. 
Since ${\cal R}_sQ^{(2)}{\cal R}_s^{-1} \!=\! -\bar{Q}^{(2)}$, 
the SU(3) and $\bsu3$ DS spectra are identical and 
the quadrupole moments of corresponding states differ in sign. 

Following the procedure of Eq.~(\ref{bases12}), 
the intrinsic part of the critical-point Hamiltonian, 
relevant to prolate-oblate (P-O) coexistence, is required to~satisfy
\bsub
\ba
\hat{H}\ket{N,\, (\lambda,\mu)=(2N,0),\,K=0,\, L} &=& 0 ~,
\label{2N0p}
\\
\hat{H}\ket{N,\, (\blam,\bmu)=(0,2N),\,\bar{K}=0,\, L} &=& 0 ~.
\label{02N}
\ea
\label{HvanishPO}
\esub 
Equivalently, $\hat{H}$ annihilates the intrinsic states of 
Eq.~(\ref{int-state}), with $(\beta\!=\!\sqrt{2},\gamma\!=\!0)$ and 
$(\beta\!=\!-\sqrt{2},\gamma\!=\!0)$, which are the lowest- and 
highest-weight vectors in the irreps $(2N,0)$ and $(0,2N)$ 
of SU(3) and $\bsu3$, respectively. 
The resulting intrinsic Hamiltonian is found to be~\cite{LevDek16},
\ba
\hat{H} = 
h_0\,P^{\dag}_0\hat{n}_sP_0 + h_2\,P^{\dag}_0\hat{n}_dP_0 
+\eta_3\,G^{\dag}_3\cdot\tilde{G}_3 ~,
\label{HintPO}
\ea
where $G^{\dag}_{3m} \!=\! \sqrt{7}[(d^{\dag} d^{\dag})^{(2)}d^{\dag}]^{(3)}_{m}
\!=\! (d^{\dag} P^{\dag}_2)^{(3)}_m \!=\! (d^{\dag} \bar{P}^{\dag}_2)^{(3)}_m$
and $P^{\dag}_{0},\,P^{\dag}_{2m},\,\bar{P}^{\dag}_{2m}$, are defined 
in Eqs.~(\ref{P0P2}) 
and (\ref{Pbar2}). The corresponding energy surface, 
$E_{N}(\beta,\gamma) = N(N-1)(N-2)\tilde{E}(\beta,\gamma)$, 
is given by
\ba
\tilde{E}(\beta,\gamma) = 
\left\{(\beta^2-2)^2
\left [h_0 + h_2\beta^2\right ] 
+\eta_3 \beta^6(1-\Gamma^2)\right \}(1+\beta^2)^{-3} ~.
\label{surfacePO}
\ea
\begin{figure}[t]
  \centering
\includegraphics[width=16cm]{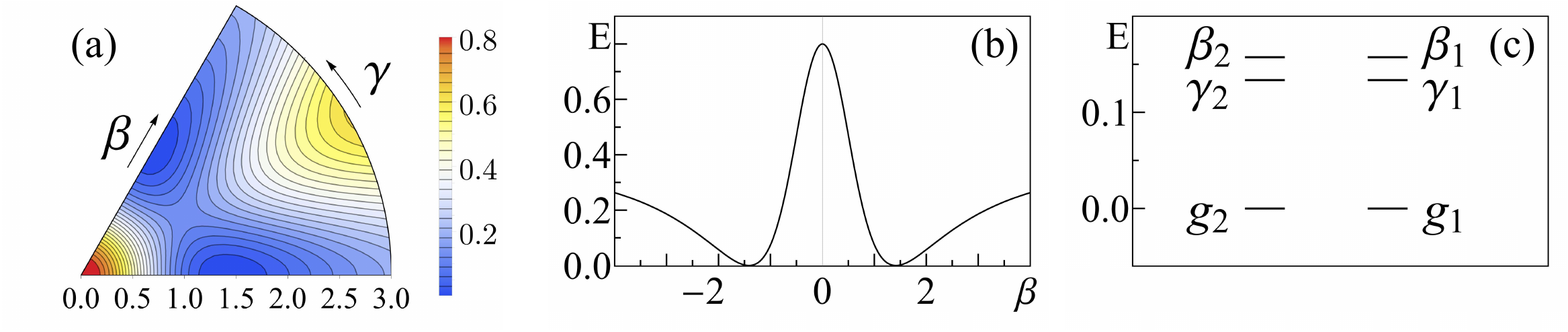}
\caption{\label{fig6-PO}
\small
Prolate-oblate (P-O) shape coexistence.
(a)~Contour plots of the energy surface~(\ref{surfacePO}),   
(b)~$\gamma\!=\!0$ sections, and
(c)~bandhead spectrum, for the Hamiltonian $\hat{H}'$~(\ref{HprimePO})
with parameters $h_0\!=\!0.2,\,h_2\!=\!0.4,\,\eta_3\!=\!0.571,
\,\alpha\!=\!0.018,\,\rho=0$ and $N\!=\!20$.} 
\end{figure}
\begin{figure}[t]
\centering
\includegraphics[width=0.9\linewidth]{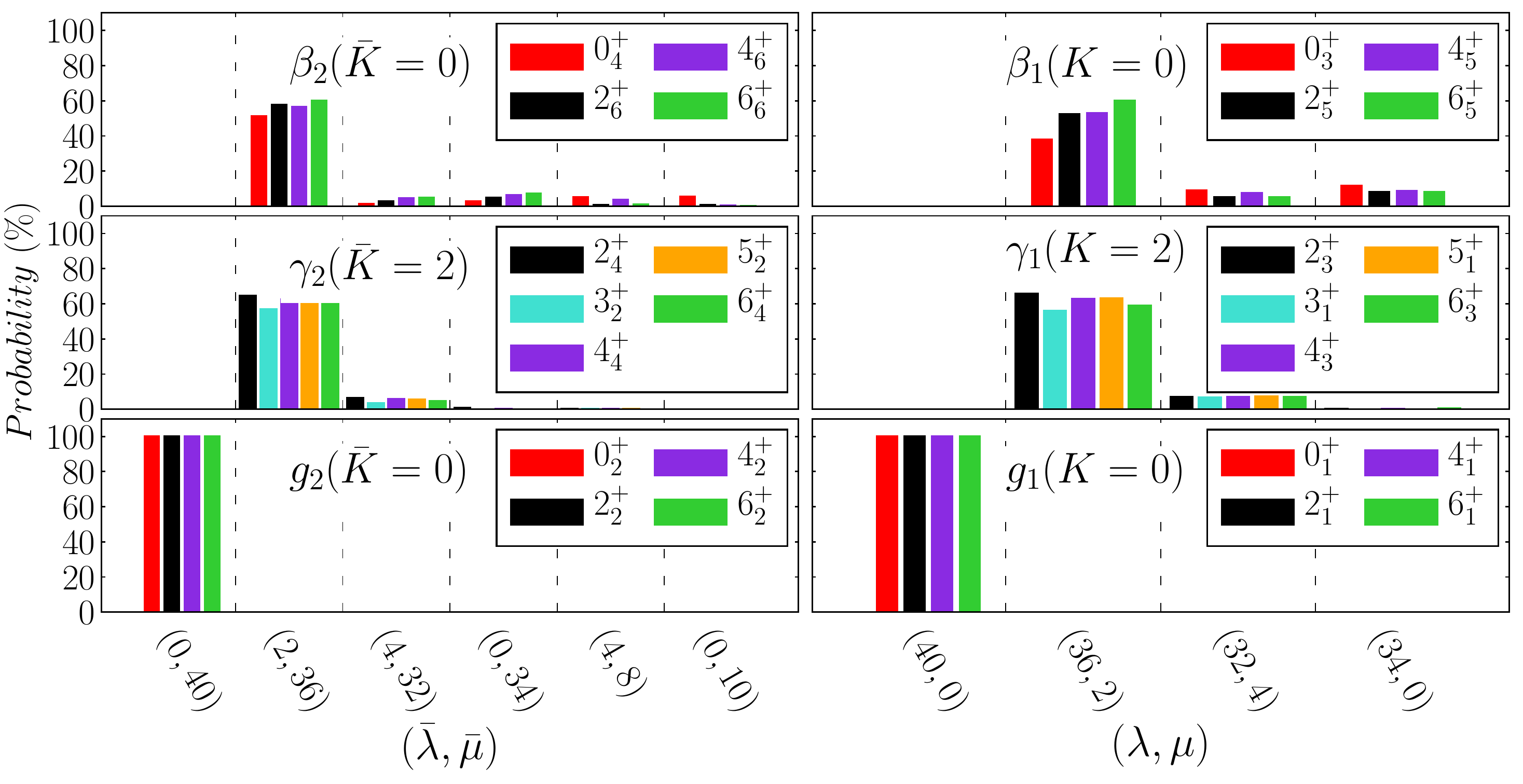}
\caption{\label{fig7-PO}
\small
SU(3) $(\lambda,\mu)$- and $\bsu3$ $(\blam,\bmu)$-decompositions 
for members of the prolate ($g_1,\beta_1,\gamma_1$) 
and oblate ($g_2,\beta_2,\gamma_2$) bands, eigenstates of 
$\hat{H}'$ (\ref{HprimePO}) with parameters as in Fig.~6, 
resulting in prolate-oblate (P-O) shape coexistence. 
Shown are probabilities larger than 5\%.}
\end{figure}
The surface is an even function of $\beta$ and 
$\Gamma = \cos 3\gamma$. For $h_0,h_2,\eta_3\geq 0$, 
$\hat{H}$ is positive definite and 
$\tilde{E}(\beta,\gamma)$ has two degenerate global minima, 
$(\beta\!=\!\sqrt{2},\gamma\!=\!0)$ and 
$(\beta\!=\!\sqrt{2},\gamma\!=\!\pi/3)$ 
[or equivalently $(\beta\!=\!-\sqrt{2},\gamma\!=\!0)$], at $\tilde{E}=0$.
$\beta=0$ is always an extremum, which is a local minimum (maximum) for 
$h_2 \!-\! 4h_0>0$ ($h_2 \!-\! 4h_0<0$), at $\tilde{E}=4h_0$.
Additional extremal points include saddle points at 
$[\beta_1\!>\!0,\gamma\!=\!0,\pi/3]$, $[\beta_2\!>\!0,\gamma\!=\!\pi/6]$ 
and a local maximum at $[\beta_3\!>\!0,\gamma\!=\!\pi/6]$. 
The saddle points, when exist, support 
a barrier separating the various minima, as seen in Fig.~6. 
For large $N$, the normal modes involve 
$\beta$ and $\gamma$ vibrations about the 
respective deformed minima, with frequencies
\bsub
\ba
&&
\epsilon_{\beta 1}=\epsilon_{\beta 2} 
= \frac{8}{3}(h_0+ 2h_2)N^2 ~,
\\
&&\epsilon_{\gamma 1}=\epsilon_{\gamma 2} = 4\eta_3N^2 ~.
\ea
\label{d-modes}
\esub
The members of the prolate and oblate ground-bands, 
Eq.~(\ref{HvanishPO}), 
are zero-energy eigenstates of $\hat{H}$ (\ref{HintPO}), 
with good SU(3) and $\bsu3$ symmetry, respectively. 
The Hamiltonian is invariant under a change of sign 
of the $s$-bosons, hence commutes with the ${\cal R}_{s}$ operator 
mentioned above. 
Consequently, all non-degenerate eigenstates of $\hat{H}$ 
have well-defined $s$-parity. 
This implies vanishing quadrupole moments for the $E2$ 
operator~(\ref{TE2}), which is odd under such sign change.
To overcome this difficulty, we introduce a small $s$-parity 
breaking term, ${\textstyle\alpha\hat{\theta}_2}$~(\ref{theta2}), 
which contributes to $\tilde{E}(\beta,\gamma)$ a component 
${\textstyle\tilde{\alpha}(1+\beta^2)^{-2}[ 
(\beta^2\!-\!2)^2 \!+\! 2\beta^2(2 \!-\!2\sqrt{2}\beta\Gamma 
\!+\!\beta^2)]}$, 
with ${\textstyle\tilde{\alpha}=\alpha/(N-2)}$. 
The linear $\Gamma$-dependence distinguishes 
the two deformed minima and slightly lifts 
their degeneracy, as well as that of the normal modes~(\ref{d-modes}). 
Identifying the collective part with $\hat{C}_2[{\rm SO(3)}]$, 
we arrive at the following complete Hamiltonian 
\ba
\hat{H}' &=& 
h_0\,P^{\dag}_0\hat{n}_sP_0 + h_2\,P^{\dag}_0\hat{n}_dP_0 
+\eta_3\,G^{\dag}_3\cdot\tilde{G}_3
+ \alpha\,\hat{\theta}_2 
+ \rho\,\hat{C}_2[\rm SO(3)] ~.\qquad
\label{HprimePO}
\ea
The prolate $g_1$-band 
remains solvable with energy $E_{g1}(L)=\rho L(L+1)$.
The oblate $g_2$-band experiences a slight shift of
order ${\textstyle\tfrac{32}{9}\alpha N^2}$ and 
displays a rigid-rotor like spectrum. 
Replacing ${\textstyle\hat{\theta}_2}$ 
by $\textstyle{\hat{\bar{\theta}}_2}$~(\ref{thetab2}), 
reverses the sign of the linear $\Gamma$ term in the energy surface 
and leads to similar effects, but 
interchanges the role of prolate and oblate bands. 
The SU(3) and $\bsu3$ decompositions in Fig.~7 demonstrate 
that these bands are pure DS basis states, with 
$(2N,0)$ and $(0,2N)$ character, respectively, 
while excited $\beta$ and $\gamma$ bands exhibit considerable mixing.
The critical-point Hamiltonian thus has a subset of states with good SU(3) 
symmetry, a subset of states with good $\bsu3$ symmetry and all other states 
are mixed with respect to both SU(3) and $\bsu3$. These are precisely the 
defining ingredients of SU(3)-PDS coexisting with $\bsu3$-PDS. 

Since the wave functions for the 
members of the $g_1$ and $g_2$ bands 
are known, one can derive analytic expressions for their 
quadrupole moments and $E2$ rates. 
For the $E2$ operator of Eq.~(\ref{TE2}),
the quadrupole moments are found to have equal magnitudes 
and opposite signs, 
\ba
Q_L &=& 
{\textstyle
\mp e_B\sqrt{\frac{16\pi}{40}}\frac{L}{2L+3}
\frac{4(2N-L)(2N+L+1)}{3(2N-1)}} ~,
\label{quadmom}
\ea
where the minus (plus) sign corresponds to the prolate-$g_1$ (oblate-$g_2$) 
band. The $B(E2)$ values for intraband ($g_1\to g_1$, $g_2\to g_2$) 
transitions, 
\ba
&&B(E2; g_i,\, L+2\to g_i,\,L) = 
\nonumber\\
&&
\quad
\;\;
{\textstyle
e_{B}^2\,\frac{3(L+1)(L+2)}{2(2L+3)(2L+5)}
\frac{(4N-1)^2(2N-L)(2N+L+3)}{18(2N-1)^2}} ~,
\qquad\qquad
\label{be2}
\ea
are the same. These properties are ensured by 
${\cal R}_sT(E2){\cal R}_s^{-1} = -T(E2)$. Interband 
$(g_2\leftrightarrow g_1)$ 
$E2$ transitions, are extremely weak. This follows from the fact that 
the $L$-states of the $g_1$ and $g_2$ bands exhaust, respectively, 
the $(2N,0)$ and $(0,2N)$ irrep of SU(3) and $\bsu3$. 
$T(E2)$ contains a $(2,2)$ tensor under both algebras, 
hence can connect the $(2N,0)$ irrep of $g_1$ only with the $(2N-4,2)$ 
component in $g_2$, which
is vanishingly small. The selection rule $g_2\nleftrightarrow g_1$ 
is valid also for a more general $E2$ operator, 
obtained by including in it the operators 
$Q^{(2)}$ or $\bar{Q}^{(2)}$, since the latter, as generators, 
cannot mix different irreps of SU(3) or $\bsu3$. 
By similar arguments, $E0$ transitions in-between 
the $g_1$ and $g_2$ bands are extremely weak, 
since the relevant operator, 
$T(E0)\propto\hat{n}_d$, is a combination of $(0,0)$ and $(2,2)$ 
tensors under both algebras. 
In contrast to $g_1$ and $g_2$, excited 
$\beta$ and $\gamma$ bands are mixed, hence are connected by 
$E2$ transitions to these ground bands. 
\begin{figure}[t]
\centering
\includegraphics[width=0.35\linewidth]{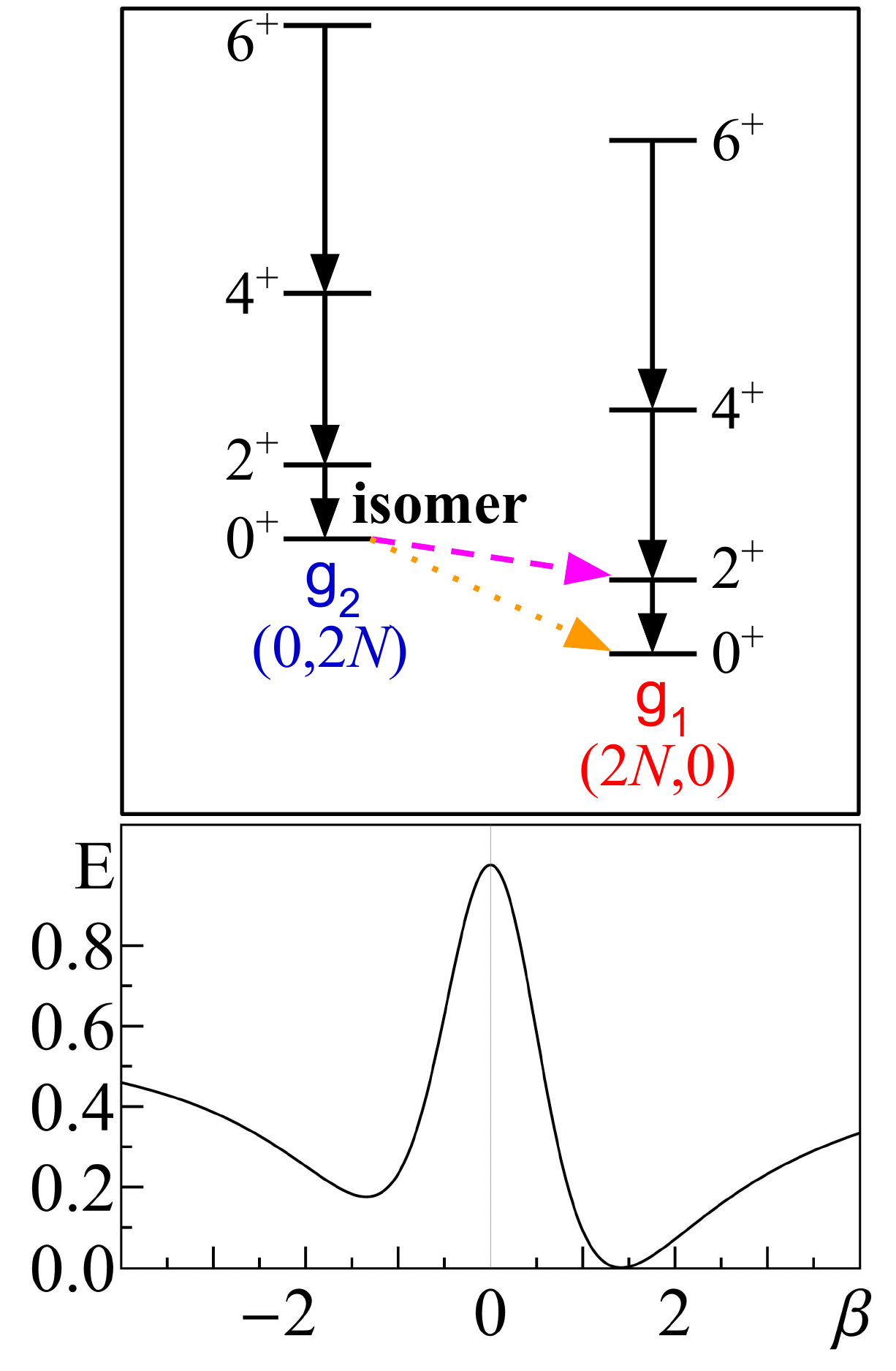}
\hspace{0.5cm}%
\includegraphics[width=0.35\linewidth]{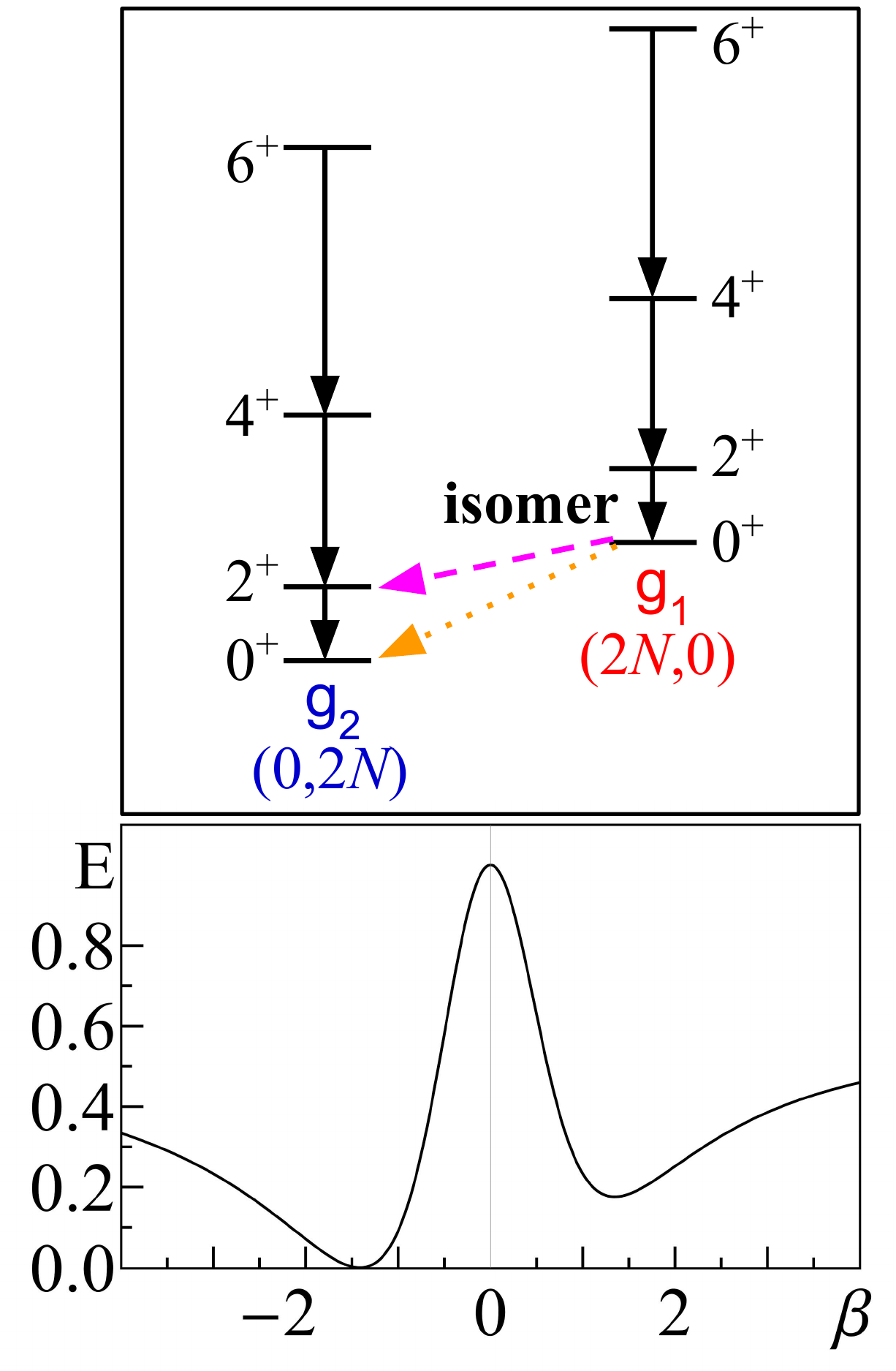}
\caption{\label{fig8-PO}
\small
Energy-surface sections and level schemes 
corresponding to departures from the critical point of P-O 
shape coexistence, for $\hat{H}'$~(\ref{HprimePO}) with parameters 
$h_0,\,h_2,\,\eta_3$ as in Fig.~6.
Left panels: an oblate isomeric state 
($\alpha\hat{\theta}_2$ term~(\ref{theta2}), with $\alpha\!=\!0.9$, 
added to $\hat{H}'$). 
Right panels: a prolate isomeric state 
($\bar{\alpha}\hat{\bar{\theta}}_2$ term~(\ref{thetab2}), 
with $\bar{\alpha}\!=\!0.9$, added to $\hat{H}'$).
Retarded $E2$ (dashed lines) and $E0$ (dotted lines) decays identify 
the isomeric states.}
\end{figure}

Departures from the critical point, can be studied by 
varying the coupling constant of the $\alpha\hat{\theta}_2$ 
term~(\ref{theta2}) in $\hat{H}'$~(\ref{HprimePO}). 
Taking larger values of $\alpha$, will leave 
the prolate $g_1$-band unchanged, but will shift the oblate $g_2$-band 
to higher energy of order $16\alpha N^2/9$. 
Similar effects are obtained by varying the strength of the 
$\bar{\alpha}\hat{\bar{\theta}}_2$ term~(\ref{thetab2}), but 
now the role of $g_1$ and $g_2$ is interchanged.
The resulting topology of the energy surfaces with such modifications 
are shown at the bottom row of Fig.~8. If these departures 
from the critical points are small, the results of Fig.~7, 
Eqs.~(\ref{quadmom})-(\ref{be2}) and the selection rules remain 
valid to a good approximation. 
In such a case, the $L=0$ bandhead state of the higher $g_i$-band 
cannot decay by strong $E2$ or $E0$ transitions to the lower ground band, 
hence, as depicted schematically on the top row of Fig.~8, 
displays characteristic features of an isomeric state.
\begin{figure}[t!]
\centering
\includegraphics[width=16cm]{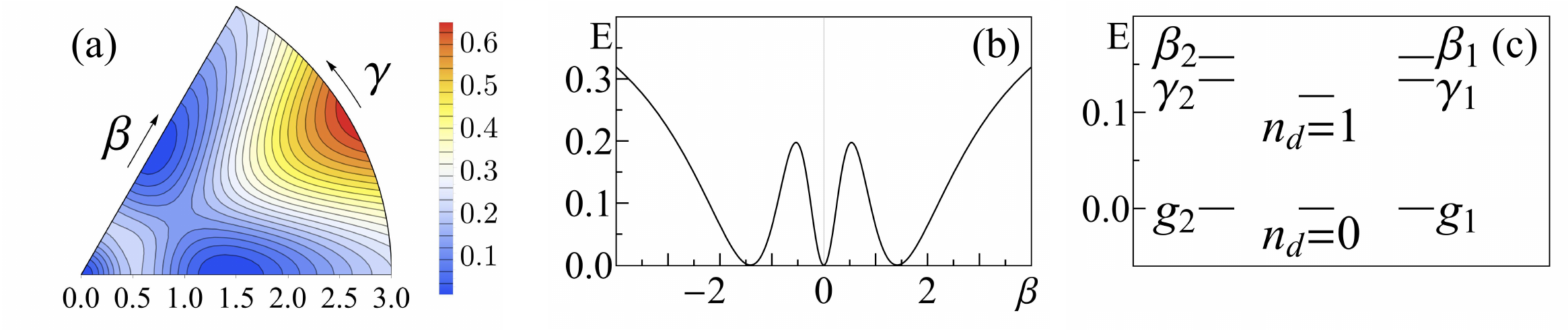}
\caption{\label{fig9-SPO}
\small
Spherical-prolate-oblate (S-P-O) shape coexistence.
(a)~Contour plots of the energy surface~(\ref{surfaceSPO}),   
(b)~$\gamma\!=\!0$ sections, and
(c)~bandhead spectrum, for the Hamiltonian $\hat{H}'$~(\ref{HprimeSPO})
with parameters $h_2\!=\!0.5,\,\eta_3\!=\!0.571,\,\alpha\!=\!0.018,
\,\rho=0$ and $N\!=\!20$.}
\end{figure}

\section{Simultaneous occurrence of spherical and two deformed shapes}

Nuclei can accommodate more than two shapes simultaneously.
A notable example is the observed coexistence of spherical, 
prolate and oblate shapes in $^{186}$Pb~\cite{Andreyev00}. 
In what follows, we consider a case study of PDS Hamiltonians relevant 
to such triple coexistence of a spherical shape 
($\beta=0$) and two axially-deformed shapes with $\gamma=0,\pi/3$ and 
equal $\beta$ deformations.

\subsection{Coexisting U(5)-PDS, SU(3)-PDS and $\bsu3$-PDS} 
\label{SPOshapes}

The DS limits relevant for spherical, prolate-deformed and 
oblate-deformed shapes, correspond to the 
chains~(\ref{U5-ds}), (\ref{SU3-ds}) and (\ref{SU3bar-ds}), respectively. 
The intrinsic part of the critical-point Hamiltonian for triple coexistence 
of such shapes is now required to satisfy three conditions
\bsub
\ba
\hat{H}\ket{N,\, n_d=0,\,\tau=0,\,L=0} &=& 0 ~,
\label{nd0tri}
\\
\hat{H}\ket{N,\, (\lambda,\mu)=(2N,0),\,K=0,\, L} &=& 0 ~,
\label{2N0tri}
\\
\hat{H}\ket{N,\, (\blam,\bmu)=(0,2N),\,\bar{K}=0,\, L} &=& 0 ~.
\label{02Ntri}
\ea
\label{vanishtri}
\esub
Equivalently, $\hat{H}$ annihilates the 
spherical intrinsic state of 
Eq.~(\ref{int-state}) with $\beta=0$, 
which is the single basis state in the U(5) irrep $n_d=0$, and 
the deformed intrinsic states with $(\beta\!=\!\sqrt{2},\gamma\!=\!0)$ 
and $(\beta\!=\!-\sqrt{2},\gamma\!=\!0)$, which are the lowest and 
highest-weight vectors in the irreps $(2N,0)$ and $(0,2N)$ 
of SU(3) and $\bsu3$, respectively. 
The resulting intrinsic Hamiltonian is found to be that of 
Eq.~(\ref{HintPO}) with $h_0=0$~\cite{LevDek16}, 
\ba
\hat{H} = 
h_2\,P^{\dag}_0\hat{n}_dP_0 +\eta_3\,G^{\dag}_3\cdot\tilde{G}_3 ~.
\label{HintSPO}
\ea
The corresponding energy surface,
\ba
\tilde{E}(\beta,\gamma) = 
\left [h_2\beta^2(\beta^2-2)^2 
+\eta_3 \beta^6(1-\Gamma^2)\right ](1+\beta^2)^{-3} ~,
\label{surfaceSPO}
\ea
has now three degenerate global minima: 
$\beta\!=\!0$, $(\beta\!=\!\sqrt{2},\gamma\!=\!0)$ 
and $(\beta\!=\!\sqrt{2},\gamma\!=\!\pi/3)$ 
[or equivalently $(\beta\!=\!-\sqrt{2},\gamma\!=\!0)$], 
at $\tilde{E}\!=\!0$, separated by barriers as seen in Fig.~9. 
In addition to the deformed $\beta$- and $\gamma$ modes of 
Eq.~(\ref{d-modes}) with $h_0=0$, there are now also spherical modes, 
involving quadrupole vibrations about the spherical minimum, 
with frequency
\ba
\epsilon = 4h_2N^2 ~.
\label{s-modes}
\ea
For the same arguments as in the analysis of P-O coexistence 
in Section~4, the complete Hamiltonian is taken to be 
that of Eq.~(\ref{HprimePO}) with $h_0=0$,
\ba
\hat{H}' &=& 
h_2\,P^{\dag}_0\hat{n}_dP_0 +\eta_3\,G^{\dag}_3\cdot\tilde{G}_3 
+ \alpha\,\hat{\theta}_2 
+ \rho\,\hat{C}_2[\rm SO(3)] ~.
\label{HprimeSPO}
\ea
\begin{figure}[t]
  \centering
\includegraphics[width=0.9\linewidth]{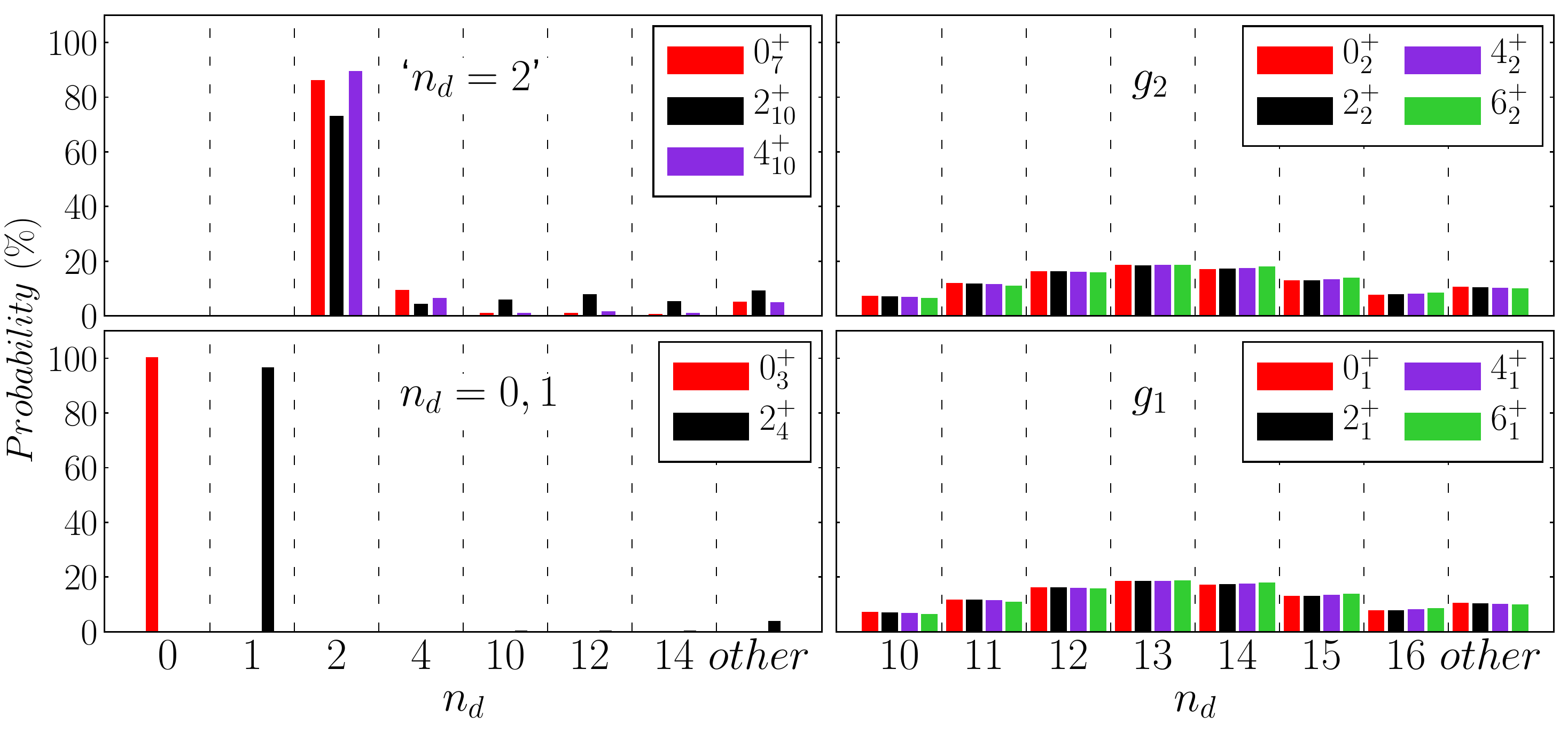}
\caption{\label{fig10-SPO}
\small
U(5) $n_d$-decomposition for spherical states (left panels)
and for members of the deformed prolate ($g_1$) and oblate ($g_2$) 
ground bands (right panels), 
eigenstates of 
$\hat{H}'$~(\ref{HprimeSPO}) with parameters as in Fig.~9, 
resulting in spherical-prolate-oblate (S-P-O) shape coexistence.
The column `other' depicts a sum of probabilities, each less than 5\%.} 
\end{figure}
The deformed bands show similar rigid-rotor structure 
as in the P-O case. 
In particular, the prolate $g_1$-band and oblate $g_2$-band have 
good SU(3) and $\bsu3$ symmetry, respectively, 
while excited $\beta$ and $\gamma$ bands 
exhibit considerable mixing, with similar decompositions as in Fig.~7.
A new aspect in the present S-P-O analysis, 
is the simultaneous occurrence in the spectrum [see Fig.~9(c)] 
of spherical type of states, whose wave functions are dominated by 
a single $n_d$ component. 
As shown in Fig.~10, the lowest spherical states have quantum numbers 
$(n_d\!=\!L\!=\!0)$ and $(n_d\!=\!1,L\!=\!2)$, 
hence coincide with pure U(5) basis states, while higher spherical states 
have a pronounced ($\sim$70\%) $n_d\!=\!2$ component. This structure 
should be contrasted with the U(5) decomposition of deformed states 
(belonging to the $g_1$ and $g_2$ bands) which, as shown in 
Fig.~10, have a broad $n_d$-distribution.
The purity of selected sets of states 
with respect to SU(3), $\bsu3$ and U(5), in the presence of other 
mixed states, are the hallmarks of 
coexisting partial dynamical symmetries.
\begin{figure}[t]
\centering
\includegraphics[width=0.35\linewidth]{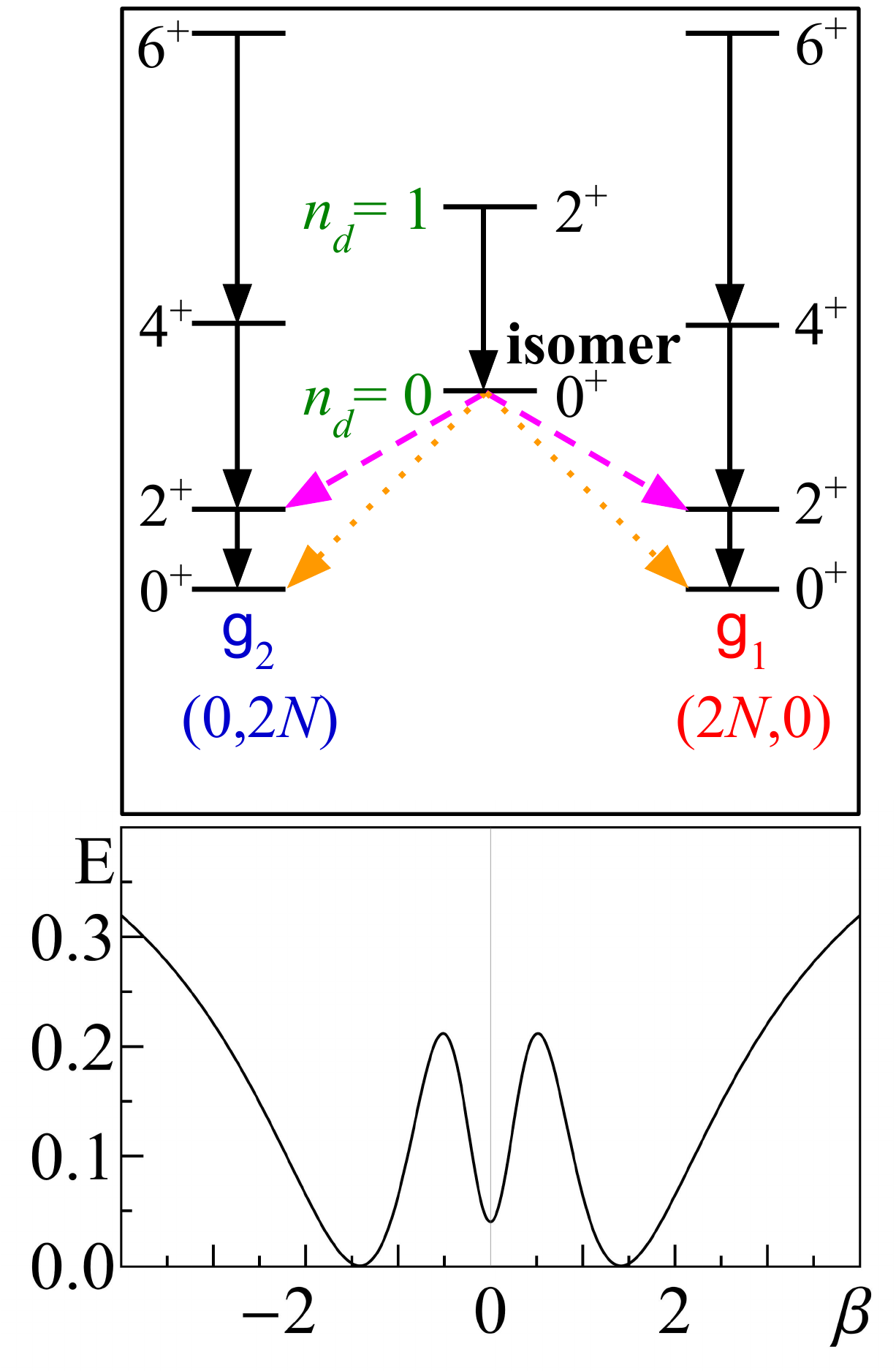}
\hspace{0.5cm}%
\includegraphics[width=0.35\linewidth]{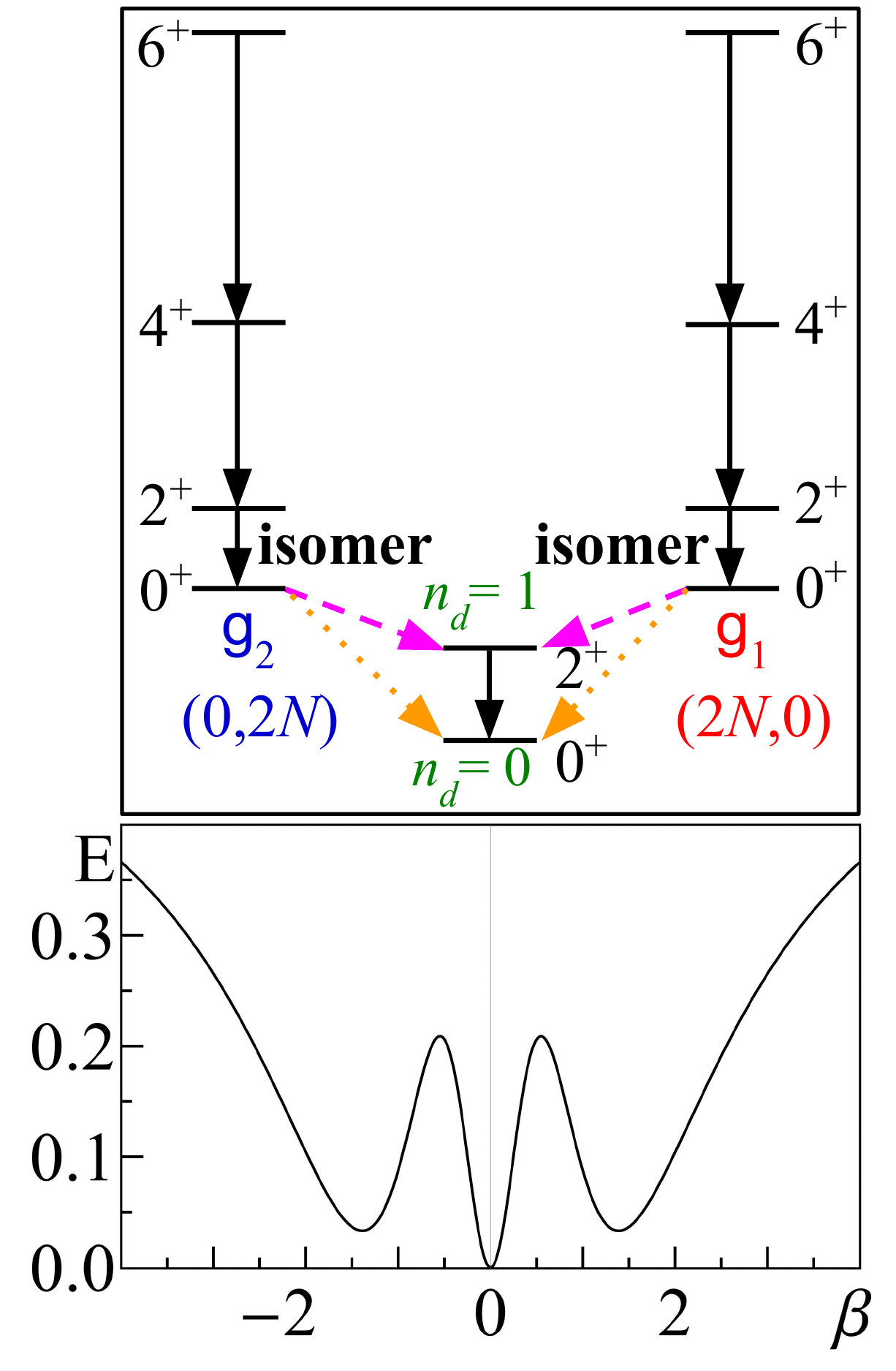}
\caption{\label{fig11-SPO}
\small
Energy-surface sections and level schemes 
corresponding to departures from the critical point of S-P-O 
shape coexistence, for $\hat{H}'$~(\ref{HprimeSPO}) 
with $h_2\!=\!0.5,\,\eta_3\!=\!0.571,\,\alpha\!=\!0.018$ 
and $N\!=\!20$. 
Left panels: a spherical isomeric state 
($h_0P^{\dag}P_0$ term~(\ref{HintPO}), 
with $h_0\!=\!0.01$, added to $\hat{H}'$). 
Right panels: P-O deformed isomeric states 
($\epsilon\hat{n}_d$ term, with $\epsilon\!=\!10$, 
added to $\hat{H}'$). 
Retarded $E2$ (dashes lines) and $E0$ (dotted lines) decays identify 
the isomeric states.}
\end{figure}

For the $E2$ operator of Eq.~(\ref{TE2}), the quadrupole moments of 
states in the solvable $g_1$ and $g_2$ bands and 
intraband ($g_1\to g_1$, $g_2\to g_2$) $E2$ rates, 
obey the analytic expressions of 
Eqs.~(\ref{quadmom}) and (\ref{be2}), respectively. 
The same selection rules depicted in Fig.~8, resulting in retarded 
$E2$ and $E0$  
interband  ($g_2\to g_1$) decays, still hold. 
Furthermore, in the current S-P-O case, 
since $T(E2)$ obeys the selection rule $\Delta n_d\!=\!\pm 1$, the 
spherical states, $(n_d\!=\!L\!=\!0)$ and $(n_d\!=\!1,L\!=\!2)$, 
have no quadrupole moment and the $B(E2)$ value for their 
connecting transition, obeys the U(5)-DS expression of Eq.~(\ref{be2nd}).
These spherical states have very weak $E2$ transitions to the 
deformed ground bands, because they exhaust the $(n_d\!=\!0,1)$ irreps 
of U(5), and the $n_d\!=\!2$ component in the ($L\!=\!0,2,4$) states 
of the $g_1$ and $g_2$ bands is 
extremely small, of order $N^33^{-N}$. 
There are also no $E0$ transitions involving these spherical states, 
since $T(E0)$ is diagonal in $n_d$.

In the above analysis, the spherical and deformed minima were assumed 
to be degenerate. If the spherical minimum is only 
local, one can use the Hamiltonian of Eq.~(\ref{HprimePO}) 
with the condition $h_2>4h_0$, 
for which the spherical ground state $(n_d=L=0)$ 
experiences a shift of order $4h_0N^3$.
Similarly, if the deformed minima are only local, 
adding an $\epsilon\hat{n}_d$ term to $\hat{H}'$ (\ref{HprimeSPO}), 
will leave the $n_d=0$ spherical ground state unchanged, but will shift 
the prolate and oblate bands to higher energy of order $2\epsilon N/3$. 
In both scenarios, the lowest $L=0$ state of the non-yrast configuration 
will exhibit retarded $E2$ and $E0$ decays, hence will have the character 
of an isomer state, as depicted schematically in Fig.~11.

\section{Concluding remarks}

We have presented an algebraic symmetry-based approach for describing 
properties of multiple shapes in dynamical systems. 
The main ingredients of the approach are: 
(i)~a spectrum generating algebra encompassing 
several lattices of dynamical symmetry (DS) chains. 
(ii)~An associated geometric space, realized by means of coherent states,
which assign a particular shape to a given DS chain. 
(iii)~An intrinsic-collective resolution of the Hamiltonian. 
The approach involves the construction of a single number-conserving, 
rotational-invariant Hamiltonian which captures the essential 
features of the dynamics near the critical point, 
where two (or more) shapes coexist.
The Hamiltonian conserves the dynamical symmetry (DS) for 
selected bands of states, associated with each shape.
Since different structural phases correspond to incompatible 
(non-commuting) dynamical symmetries, the symmetries in question 
are shared by only a subset of states, and are broken in the remaining 
eigenstates of the Hamiltonian. 
The resulting structure is, therefore, that of coexisting multiple 
partial dynamical symmetries (PDSs). 

An explicit algorithm for constructing Hamiltonians with several 
distinct PDS was presented and the approach was applied to a variety 
of coexisting quadrupole shapes, 
in the framework of the interacting boson model (IBM) of nuclei. 
The multiple PDSs and shape-coexistence scenarios considered include 
(i)~Coexisting U(5) and SU(3) PDSs, 
in relation to spherical and prolate-deformed shapes. 
(ii)~Coexisting U(5) and $\bsu3$ PDSs, 
in relation to spherical and oblate-deformed shapes. 
(iii)~Coexisting U(5) and SO(6) PDSs, 
in relation to spherical and $\gamma$-unstable deformed shapes.
(iv)~Coexisting SU(3) and $\bsu3$ PDSs, 
in relation to prolate and oblate deformed shapes.
(v)~Coexisting U(5), SU(3) and $\bsu3$ PDSs, 
in relation to spherical, prolate-deformed and oblate-deformed shapes.

In each of the cases considered, the underlying energy surface exhibits 
multiple minima which are near degenerate. 
As shown, the constructed Hamiltonian has the capacity to have 
distinct families of states whose properties reflect the different 
nature of the coexisting shapes. 
Selected sets of states within each family, 
retain the dynamical symmetry associated with the given shape. 
This allows one to obtain 
closed expressions for quadrupole moments and transition 
rates, which are the observables most closely related to the nuclear shape.  
The resulting analytic expressions 
are parameter-free predictions, except for a scale, and can be used 
to compare with measured values of these observables 
and to test the underlying partial symmetries.
The purity and good quantum numbers of selected states 
enable the derivation of symmetry-based selection rules for 
electromagnetic transitions (notably, for $E2$ and $E0$ decays) and the 
subsequent identification of isomeric states. 
The evolution of structure away from the critical-point 
can be studied by adding to the Hamiltonian the Casimir operator 
of a particular DS chain, which will leave unchanged the ground band of 
one configuration but will shift the other configuration(s) 
to higher energy.

A detailed microscopic interpretation of shape coexistence in nuclei is 
a formidable task and computational demanding~\cite{Heyde11}. 
The proposed algebraic approach presents a simple alternative, 
as a starting point to describe this phenomena, 
by emphasizing the role of remaining underlying symmetries, 
which provide physical insight and make the problem tractable. 
It is gratifying to note that shape-coexistence in dynamical systems, 
such as nuclei, constitutes a fertile ground for the development 
and testing of generalized notions of symmetry.

\ack
This work is supported by the Israel Science Foundation (Grant 586/16).


\medskip
\begin{thebibliography}{99}

\bibitem{BNB}
Bohm A, N\' eeman Y and Barut A O eds 1988 
{\it Dynamical Groups and Spectrum Generating Algebras} 
(Singapore: World Scientific)

\bibitem{ibm}
Iachello F and Arima A 1987
{\it The Interacting Boson Model}
(Cambridge University Press, Cambridge)

\bibitem{ibfm}
Iachello F and Van Isacker P 1991 
{\it The Interacting Boson-Fermion Model} 
(Cambridge: Cambridge Univ. Press)

\bibitem{vibron}
Iachello F and Levine R D 1994 
{\it Algebraic Theory of Molecules} 
(Oxford: Oxford Univ. Press)

\bibitem{LieAlg}
Iachello F 2015
{\it Lie Algebras and Applications}
(Berlin Heidelberg: Springer-Verlag)

\bibitem{gino80}
Ginocchio J N and Kirson M W 1980 
{\it Phys. Rev. Lett.} {\bf 44} 1744

\bibitem{diep80}
Dieperink A E L, Scholten O and Iachello F 1980 
{\it Phys. Rev. Lett.} {\bf 44} 1747

\bibitem{Heyde11}
Heyde K and Wood J L (2011)
{\em Rev. Mod. Phys.} {\bf 83} 1467

\bibitem{Leviatan11}
Leviatan A 2011
{\it Prog. Part. Nucl. Phys.} {\bf 66} 93

\bibitem{Alhassid92}
Alhassid Y and Leviatan A 1992
{\it J. Phys.} A {\bf 25} L1265

\bibitem{Leviatan96}
Leviatan A 1996
{\it Phys. Rev. Lett.} {\bf 77} 818

\bibitem{LevSin99}
Leviatan A and Sinai I 1999
{\it Phys. Rev.} C {\bf 60} 061301(R)

\bibitem{Casten14}
Casten R F, Cakirli R B, Blaum K, and Couture A 2014 
{\it Phys. Rev. Lett.} {\bf 113} 112501

\bibitem{Couture15}
Couture A, Casten R F, and Cakirli R B 2015
{\it Phys. Rev.} C {\bf 91} 014312

\bibitem{Casten16}
Casten R F, Jolie J, Cakirli R B, and Couture A 2016
{\it Phys. Rev.} C {\bf 94} 061303(R)

\bibitem{LevIsa02}
Leviatan A and Van Isacker P 2002
{\it Phys. Rev. Lett.} {\bf 89} 222501

\bibitem{Kremer14}
Kremer C {\it et al.} 2014 
{\it Phys. Rev.} C {\bf 89}, 041302(R); 
2015 {\it Phys. Rev.} C {\bf 92} 039902

\bibitem{LevGino00}
Leviatan A and Ginocchio J N 2000
{\it Phys. Rev.} C {\bf 61} 024305

\bibitem{GarciaRamos09}
Garc\'\i a-Ramos J E , Leviatan A and Van Isacker P 2009
{\it Phys. Rev. Lett.} {\bf 102} 112502

\bibitem{Leviatan13}
Leviatan A, Garc\'\i a-Ramos J E , and Van Isacker P 2013
{\it Phys. Rev.} C {\bf 87} 021302(R)

\bibitem{Pds-BF15}
Van Isacker P Jolie J Thomas T and Leviatan A 2015
{\it Phys. Rev.} C {\bf 92} 011301(R)

\bibitem{Isacker99}
Van Isacker P 1999
{\it Phys. Rev. Lett.} {\bf 83} 4269

\bibitem{Escher00}
Escher J and Leviatan A 2000
{\it Phys. Rev. Lett.} {\bf 84} 1866

\bibitem{Escher02}
Escher J and Leviatan A 2002
{\it Phys. Rev.} C {\bf 65} 054309

\bibitem{Rowe01}
Rowe D J and Rosensteel G 2001
{\it Phys. Rev. Lett.} {\bf 87} 172501

\bibitem{Rosen03}
Rosensteel G and Rowe D J 2003
{\it Phys. Rev. C} {\bf 67} 014303

\bibitem{Isacker08}
Van Isacker P and Heinze S 2008
{\it Phys. Rev. Lett.} {\bf 100} 052501 

\bibitem{Isacker14}
Van Isacker P and Heinze S 2014
{\it Ann. Phys. (N.Y.)} {\bf 349} 73

\bibitem{Leviatan07}
Leviatan A 2007
{\it Phys. Rev. Lett.} {\bf 98} 242502

\bibitem{Macek14}
Macek M and Leviatan A 2014
{\it Ann. Phys. (N.Y.)} {\bf 351} 302

\bibitem{LevDek16}
Leviatan A and Shapira D 2016
{\it Phys. Rev.} C {\bf 93} 051302(R)

\bibitem{LevGav17}
Leviatan A and Gavrielov N 2017
{\it Phys. Scr.} {\bf 92} 114005

\bibitem{WAL93}
Whelan N, Alhassid Y and Leviatan A 1993
{\it Phys. Rev. Lett.} {\bf 71} 2208

\bibitem{LW93}
Leviatan A and Whelan N D 1996
{\it Phys. Rev. Lett.} {\bf 77} 5202

\bibitem{kirlev85}
Kirson M W and Leviatan A 1985
{\it Phys. Rev. Lett.} {\bf 55} 2846 

\bibitem{Leviatan87}
Leviatan A 1987
{\it Ann. Phys. (N.Y.)} {\bf 179} 201

\bibitem{levkir90}
Leviatan A and Kirson M W 1990
{\it Ann. Phys. (N.Y.)} {\bf 201} 13

\bibitem{Leviatan06}
Leviatan A 2006
{\it Phys. Rev. C} {\bf 74} 051301(R) 

\bibitem{Clement16}
Cl\'ement E {\it et al.} 2016
{\it Phys. Rev. Lett.} {\bf 116} 022701

\bibitem{Park16}
Park J {\it et al.} 2016
{\it Phys. Rev.} C {\bf 93} 014315

\bibitem{kremer16}
Kremer C {\it et al.} 2016
{\it Phys. Rev. Lett.} {\bf 117} 172503

\bibitem{gottardo16}
Gottardo A {\it et al.} 2016
{\it Phys. Rev. Lett.} {\bf 116} 182501

\bibitem{yang16}
Yang X F {\it et al,} 2016
{\it Phys. Rev. Lett.} {\bf 116} 182502

\bibitem{Ayan16}
A.D. Ayangeakaa {\it et al.} 2016
{\it Phys. Lett. B} {\bf 754} 254

\bibitem{Clement07}
Cl\'ement E {\it et al.} 2007
{\it Phys. Rev.} C {\bf 75} 054313

\bibitem{Ljun08}
Ljungvall J {\it et al.} 2008
{\it Phys. Rev. Lett.} {\bf 100} 102502

\bibitem{Bree14}
Bree N {\it et al.} 2014
{\it Phys. Rev. Lett.} {\bf 112} 162701

\bibitem{Andreyev00}
Andreyev A N  {\it et al.} 2000 
{\it Nature} {\bf 405} 430

\end{thebibliography}
\end{document}